\documentclass[aps,pra,reprint, superscriptaddress,floatfix]{revtex4-1}
\usepackage{amsmath,graphicx,hyperref,xcolor}
\usepackage[utf8]{inputenc}
\usepackage[T1]{fontenc}
\usepackage{natbib}
\usepackage{amsmath,graphicx,color,enumitem,hyperref}
\usepackage{lipsum}
\usepackage[export]{adjustbox}

\usepackage{graphicx}

\usepackage{float}

\newcommand{\figref}[1]{Fig.~\ref{#1}}
\begin{document}

\title{Demonstration of Fourier-domain Quantum Optical Coherence Tomography for a fast tomographic quantum imaging}

%\title{Fourier-domain quantum optical coherence tomography: Redefining precision in quantum imaging}

\author{Sylwia M. Kolenderska}
\email{skol745@aucklanduni.ac.nz}
\affiliation{Institute of Physics, Faculty of Physics, Astronomy and Informatics, Nicolaus Copernicus University in Toruń, ul. Grudziądzka 5, 87-100 Toruń, Poland}
\affiliation{School of Physical and Chemical Sciences, University of Canterbury, Christchurch, New Zealand}

\author{F. Crislane Vieira de Brito}
%\affiliation{Institute of Physics, Faculty of Physics, Astronomy and Informatics, Nicolaus Copernicus University in Toruń, ul. Grudziądzka 5, 87-100 Toruń, Poland}

\author{Piotr Kolenderski}
\affiliation{Institute of Physics, Faculty of Physics, Astronomy and Informatics, Nicolaus Copernicus University in Toruń, ul. Grudziądzka 5, 87-100 Toruń, Poland}

%\keywords{Keyword1, Keyword2, Keyword3}

\begin{abstract}
Using spectrally correlated photon pairs instead of classical laser light and coincidence detection instead of light intensity detection, Quantum Optical Coherence Tomography (Q-OCT) outperforms classical OCT in several experimental terms. It provides twice better axial resolution with the same spectral bandwidth and it is immune to even-order chromatic dispersion, including Group Velocity Dispersion responsible for the bulk of axial resolution degradation in the OCT images. Q-OCT has been performed in the time domain configuration, where one line of the two-dimensional image is acquired by axially translating the mirror in the interferometer's reference arm and measuring the coincidence rate of photons arriving at two single-photon-sensitive detectors. Although successful at producing resolution-doubled and dispersion-cancelled images, it is still relatively slow and cannot compete with its classical counterpart. Here, we experimentally demonstrate Q-OCT in a much faster Fourier-domain configuration, theoretically proposed in 2020, where the reference mirror is fixed and the joint spectrum is acquired by inserting long fibre spools in front of the detectors. We propose two joint spectrum pre-processing algorithms, aimed at compensating resolution-degrading effects within the setup. While the first one targets fibre spool dispersion, an effect specific to this configuration, the other one removes the effects leading to the weakening of even-order dispersion cancellation – the latter impossible to be mitigated in the time-domain alternative.
Being additionally contrasted with both the time-domain approach and the conventional OCT in terms of axial resolution, imaging range and multilayer-object imaging, Fourier-domain Q-OCT is shown to be a significant step forward towards a practical and competitive solution in the OCT arena.

\end{abstract}

\maketitle

\section*{Introduction}

%QOCT
Quantum Optical Coherence Tomography (Q-OCT) was first proposed in 2002 \cite{abouraddy2002quantum} and then experimentally demonstrated in 2003 by the same group \cite{nasr2003demonstration}. It is a non-classical equivalent of conventional OCT, offering two times better axial resolution for the same spectral bandwidth and the elimination of the image-degrading even-order chromatic dispersion \cite{abouraddy2002quantum,kolenderska2020quantum}.
% These two sought-after features result from the quantum features, more specifically the spectral anticorrelation \cite{kolenderska2020quantum}, of the photon pairs used for Q-OCT imaging.

\begin{figure}[t]
    \centering
    \includegraphics[width=\linewidth]{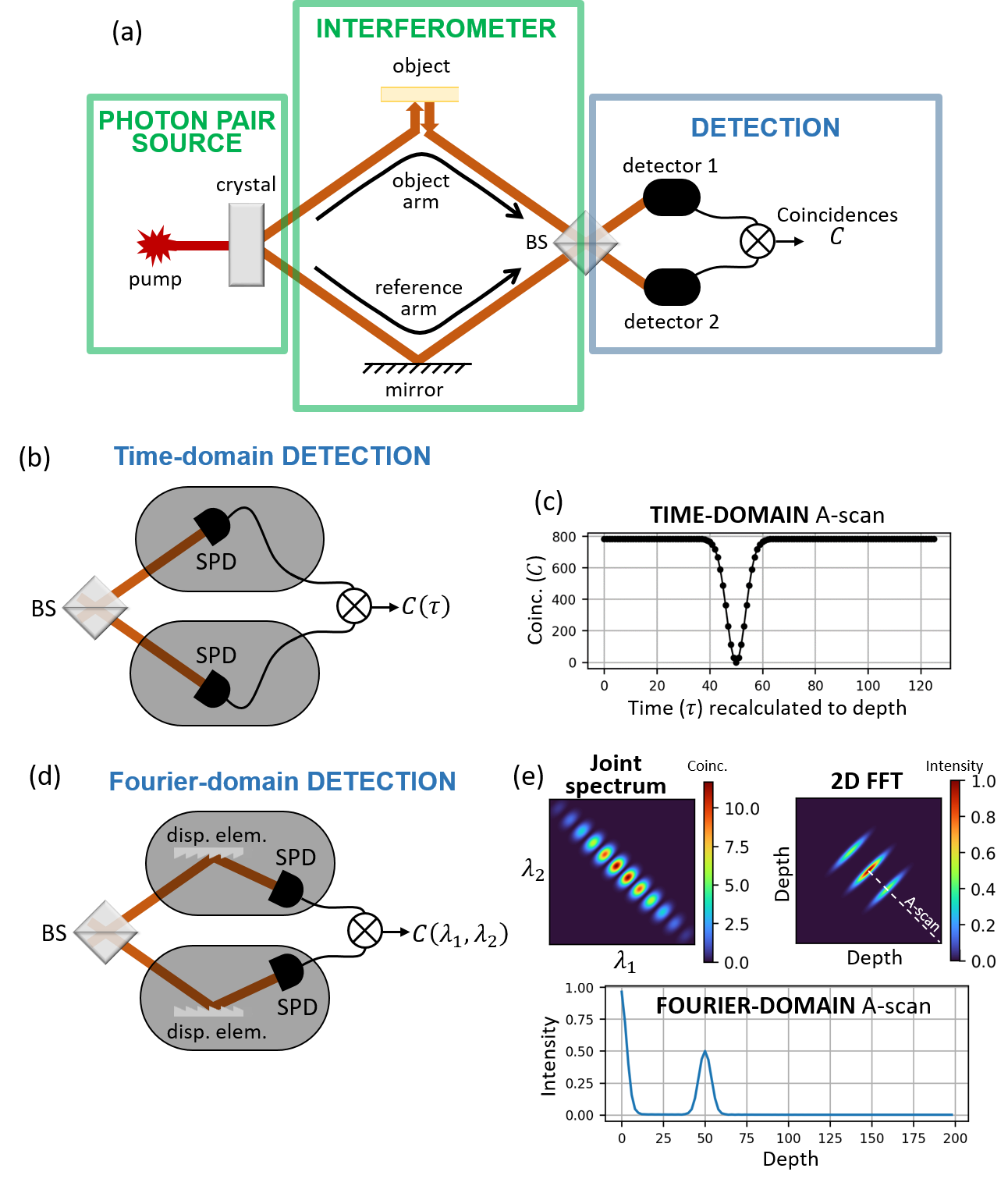}
    \caption{\protect  \textbf{Quantum OCT} comprises (a) a photon pair source, an interferometer where the produced photons propagate, and a detection part measuring coincidence \(C\) of the photons' arrival at two detectors. (b) Time-domain Q-OCT is based on single-photon-sensitive, single-pixel detectors, and outputs (c) a dip when the object arm length equals the reference arm length. (d) The Fourier-domain Q-OCT incorporates additional dispersive elements enabling wavelength discrimination, producing (e) a 2D joint spectrum which is Fourier transformed, with half the diagonal being the A-scan. The joint spectrum contains fringes along the diagonal, here, a single-frequency modulation appearing when a mirror for an object is moved away from the zero delay location and which - when Fourier transformed - produces a peak at a location proportional to the distance from the zero delay location. Such generation of different single-frequency modulations in the spectrum by different reflectors inside the object is what underpins the Fourier-domain OCT approaches.  \newline BS - beamsplitter, \(\lambda_1\), \(\lambda_2\) - wavelength of the photons in the pair, SPD - single-photon-sensitive detector.}
\label{fig:concept}
%\vspace{-20pt}
\end{figure}

 %QOCT -- simple method description
In the Q-OCT setup, conceptually presented in \figref{fig:concept}a, entangled photon pairs are generated in a nonlinear crystal and propagated in a Mach-Zehnder-type interferometer, called Hong-Ou-Mandel (HOM) interferometer \cite{hong1987measurement}. 
While one of the photons is reflected from the object, the other photon propagates in the reference arm, both subsequently meeting at the beamsplitter. Each output port of the beam-splitter has a single-photon-sensitive detector enabling the measurement of coincidences (\figref{fig:concept}b), events in which the photons in the pair exit the beamsplitter using different ports and arrive at the detectors at the same time.

% dip/time-domain A-scan creation
A dip in the coincidence signal occurs (\figref{fig:concept}c) when the reference arm length becomes equal to the object arm length. In such a situation, the photons arrive at the beam-splitter at the same time and - being indistinguishable (having the same optical parameters, e.g. the polarisation, central frequency) - they undergo quantum interference and exit together through the same output port. This reduces the number of counted coincidences, providing a mechanism for detecting back-reflecting elements in the object inserted in the object arm. Such a depth structure of an object at a lateral position, called an A-scan, is mapped by varying the reference arm length and producing a dip whenever that length is matched to the length in the object arm set by an object layer.

%demonstrations
With a variable-length reference arm and single-pixel detection, Q-OCT has been realized in what is known in the OCT community as the time-domain configuration. Not only did this configuration show a twofold increase in axial resolution and chromatic dispersion cancellation for different photon pair sources and their spectral bandwidths  (800~nm \cite{hayama2022high, yepiz2022quantum, sukharenko2021birefringence, ibarra2019experimental, okano20150, lopez2012quantum, nasr2009quantum, nasr2004dispersion, nasr2003demonstration}, 1550~nm \cite{yepiz2020spectrally, graciano2019interference}), it proved to enable very useful modalities and extensions, including polarisation-sensitive quantum imaging \cite{sukharenko2021birefringence} and quantum optical coherence microscopy imaging \cite{yepiz2022quantum}.

%problems
Despite these advances, Q-OCT is still far from surpassing conventional OCT, mainly because of the method's long measurement times. A good quality time-domain signal for a mirror requires about 1~second of integration time per data point, resulting in several minutes of acquisition time \cite{yepiz2022quantum, okano20150} per image line. The acquisition of a single image line is generally shorter for low-axial-resolution signals because the depth scan does not need to be very dense to properly sample the dips.

%problems in numbers
For a piece of glass, which is inherently much less reflective than a mirror, the integration time per data point increases to about 10~seconds and the image line acquisition time to hours, again depending on the axial resolution and additionally, on the thickness of the sample \cite{yepiz2022quantum, ibarra2019experimental}. Biological imaging with Q-OCT has only been successfully performed on objects coated with reflective particles, and the obtaining of a three-dimensional image took more than 24 hours \cite{yepiz2022quantum, nasr2009quantum}.

%solutions, time-domain and towards fourier-domain
It has been noted that the acquisition time of Q-OCT could be reduced if advances in quantum technologies continue, allowing the use of brighter photon-pair sources or more sensitive and less noisy single-photon detectors \cite{teich2012variations}. Alternatively, time-optimised measurement schemes could be explored, with one such scheme already proposed in 2019 \cite{ibarra2019experimental}. Inspired by conventional OCT, the full-field approach, in which all lines of the image are acquired simultaneously, allowed a two-dimensional en face image to be obtained in 3~minutes. Another notable advancement in this area came just one year later from the same research group \cite{yepiz2020spectrally}.  Taking a leaf out of the conventional OCT book again,  U'Ren \& Cruz-Ramirez et al.~proposed in early 2020 a partial shift in the Q-OCT detection to the Fourier domain. By measuring a one-dimensional spectrum for a given reference arm length and using it to reconstruct the time-domain signal, the acquisition of a single image line for extended objects was reduced from hours to minutes.

%Fourier-domain 
We also explored the potential of Fourier-domain measurements as a more rapid detection alternative for Q-OCT. This was initially presented at the European Conference on Biomedical Optics in June 2019 \cite{kolenderska2019} and subsequently elaborated upon in a theoretical journal article published the following year  \cite{kolenderska2020fourier}. In our approach, we proposed a complete shift to the Fourier domain by measuring a whole two-dimensional joint spectrum, as opposed to its single diagonal, as is the case in the work of U'Ren \& Cruz-Ramirez et al. \cite{yepiz2020spectrally}. A joint spectrum is obtained by counting the coincidence events in the output ports of the final beamsplitter, as is done in the time-domain modality. However, in this approach, photons in each port undergo wavelength measurement(\figref{fig:concept}d) enabling the detected coincidences to be plotted as a function of two wavelengths (\figref{fig:concept}e). The resulting joint spectrum is Fourier transformed and half its diagonal is taken to produce an A-scan \cite{kolenderska2021artefact}.

Here we present an experimental setup for Fourier-domain Q-OCT, where the joint spectral detection is realised with two 5-kilometre-long fibre coils and superconducting single photon detectors. Such a Fourier-domain Q-OCT allows for a much faster depth scan without any moving elements in the reference arm. Using a spectrally broadband laser as the pump, its image-scrambling artefacts are reduced, with the potential of being completely removed if one takes advantage of the spectral relationships elucidated in the joint spectrum.
We propose a pre-processing algorithm for compensating the joint spectrum distortions caused by the fibre spool dispersion. Also, an interpolation-based algorithm is developed for compensating the resolution-degrading effect induced by the introduction of a broadband pump \cite{okano2013dispersion}, providing the very first solution to this problem.
With an improved A-scan calculation and time-optimised joint spectrum acquisition, we compare the artefact removal capabilities as well as general imaging parameters (axial resolution, imaging range) of Fourier-domain Q-OCT imaging with that of the Time-domain one. Comparison with traditional OCT outlines the problems that need to be solved for Q-OCT to become a useful alternative.

% There were methods proposed to remove artefacts, image-scrambling elements in the signal, occuring as a by-product of using the quantum light \textcolor{blue}{[***]}.

% research  intensified especially in the last decade as the quantum technologies have become a bigger and bigger focus worldwide.   

\section*{Methods}

\subsection*{The experimental setup}

\begin{figure*}[!htb]
    \centering
    \includegraphics[width=0.7\textwidth]{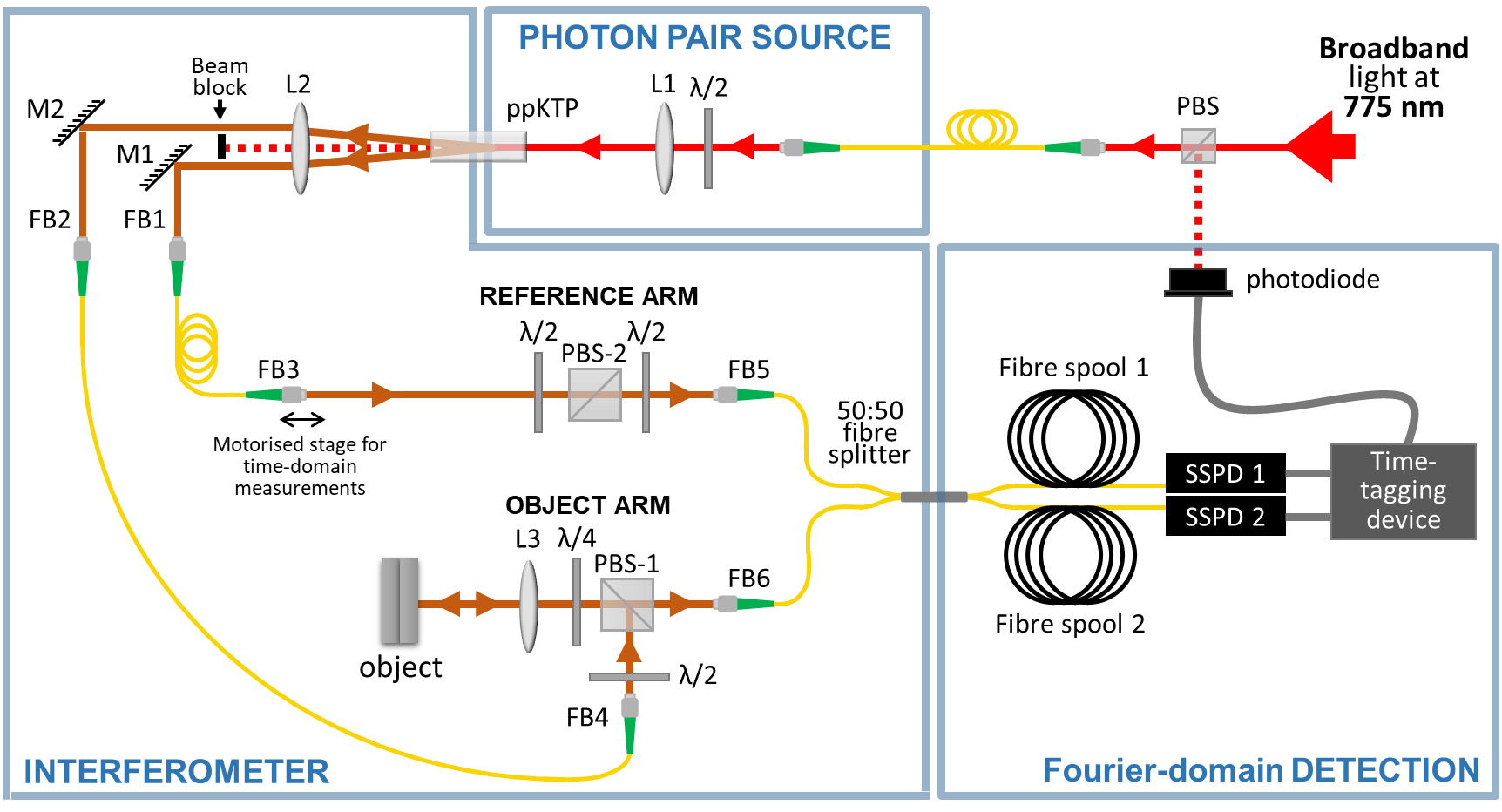}
    \caption{\textbf{Fourier-domain Quantum OCT setup} uses broadband laser light at 775~nm, Coherent Chameleon Ultra+, 80~MHz repetition rate, 10 nm bandwidth, to pump the ppKTP, 24.7~$\mu$m poling period,  crystal and generate photon pairs at 1550 nm. One photon from the pair propagates in the object arm where it is reflected from the object. The other photon from the pair propagates in the reference arm, whose length and produced polarisation match those of the object arm. The photons overlap at the 50:50 fibre splitter where they quantum interfere. The wavelength-dependent coincidence rate is measured using two fibre spools, superconducting single-photon detectors SSPDs, Scontel, and a photodiode-triggered time-tagging device, Qutag by Qutools.
    A \textbf{Time-domain~Quantum~OCT} signal is acquired by translating the motorised stage in the reference arm. 
    \textbf{Conventional OCT} is performed by connecting a second fibre splitter to a Menlo T-light laser (central wavelength of 1550~nm and the total bandwidth of 160~nm) at one end and to FB1 and FB2 at the other one. The classical fringes are then acquired using an optical spectrum analyser, Yokogawa AQ6374E, connected to one of the output ports of the 50:50 fibre splitter.  M1-2 -- mirrors, L1-3 -- lenses (f=125~mm, 75~mm, 50~mm), $\lambda$/2 -- half-wave plate, $\lambda $/4 -- quarter-wave plate, PBS - polarisation beamsplitter, fibre connectors: FB1 and FB2 -- 8-mm focal length lenses in front of HP1550 single-mode fibres, FB3 and FB4 -- collimation package, Thorlabs FC1550-f6.37~mm, FB5 and FB6 -- lens with f=11~mm in front of the fibres. }
    \label{fig:setup}
%    \vspace{-10pt}
\end{figure*}

The Fourier-domain Quantum Optical Coherence Tomography (Fd-Q-OCT) setup consists of three parts: the photon pair source,  the interferometer, and the joint spectral detection, marked by rectangles in the setup schematic in \figref{fig:setup}.

In the photon pair source, a pulsed laser light with a central wavelength of 775 nm is used as the pump for the Spontaneous Parametric Down Conversion (SPDC) process, in which entangled photon pairs are created. The pump light is focused by a lens L1 onto a ppKTP crystal, and two photons  with wavelengths around 1550~nm and polarisations identical to that of the pump photon exit the crystal in different directions, consequently realising non-collinear, type 0 SPDC. The half-wave plate situated in front of L1  adjusts the polarisation of the pump beam for the optimum pair production rate.

%In the photon pair source, a pulsed laser light with a central wavelength of 775 nm is used as the pump, and it is focused by a lens L1 onto a ppKTP crystal. In this process, non-collinear, type 0 \textcolor{red}{[Cris: I see you removed the explanation of what this type means, I still like the explanation: same pol...]} Spontaneous Parametric Down Conversion (SPDC) is employed to create the photon pairs. The half-wave plate situated in front of L1  adjusts the polarisation of the pump beam for the optimum pair production rate.

%interferometer
%The two photons in a generated pair exit the crystal at different directions, forming two arms of the interferometer. After passing the collimating lens L2 (f=75~mm), the photons are reflected by mirrors M1 and M2 and then coupled to separate fibres FB1 and FB2. While the photons input at FB1 go on to propagate in the reference arm, the photons input at FB2 go on to propagate in the object arm. In both arms, light is outputted into free space with matching light collimation packages FB3 and FB4 (Thorlabs FC1550-f6.37mm) and injected back into fibres with matching injection setups FB5 and FB6 (lens with f=11~mm).

%interferometer
The two photons  exiting the crystal in different directions form two arms of the interferometer. After passing through the collimating lens L2, the photons are reflected by mirrors M1 and M2 and then coupled into separate fibres FB1 and FB2. While the photons entering FB1 propagate in the reference arm, the photons entering FB2 propagate in the object arm. In both arms, light is output to free space with matching light collimation packages FB3 and FB4 and injected back into fibres with matching injection setups FB5 and FB6.

\begin{figure*}[t]
    \centering
    \includegraphics[width=0.9\textwidth]{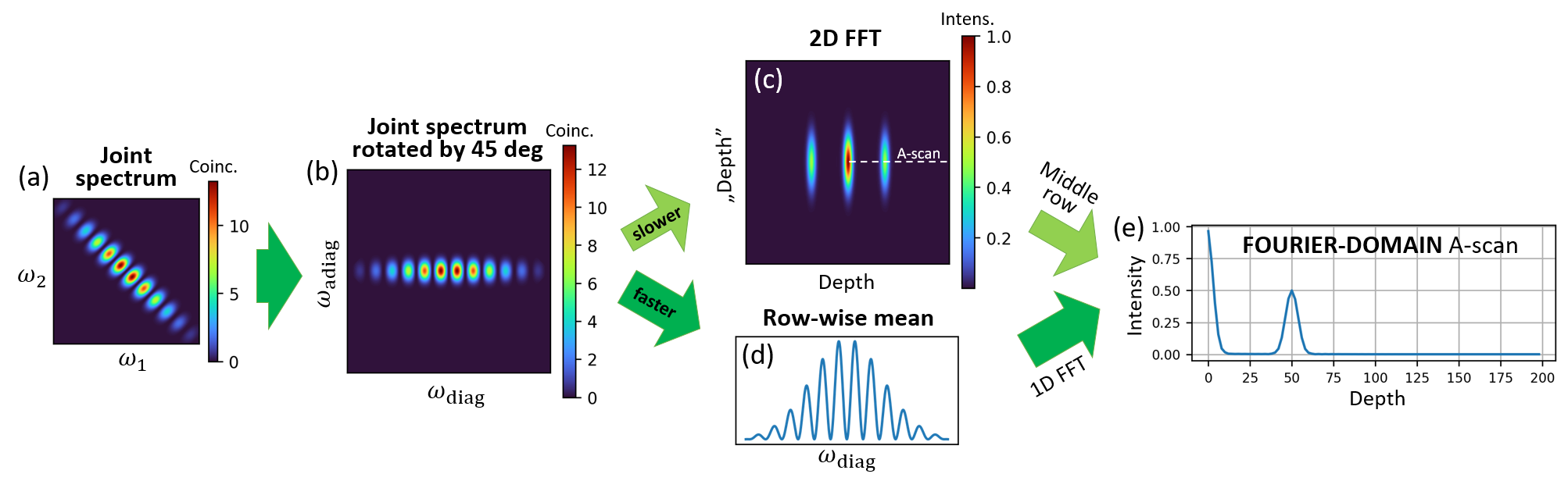}
    \caption{(a) The input joint spectrum is distributed diagonally. (b) After its counter-clockwise rotation by 45 degrees, the joint spectrum, whose signal distribution is now along the X axis, can  undergo either (c) slower 2D Fourier transformation in order to produce 2D Fourier transform, or (d) its faster row-wise mean and 1D Fourier transformation calculation, both alternative ways to obtain (e) an A-scan.  }
    \label{fig:ascan-calculation}
%    \vspace{-10pt}
\end{figure*}

In the object arm, the first half-wave plate rotates the polarisation of the light to maximise reflection at the polarising beam-splitter PBS-1, thereby directing most of the light towards the object. The L3 achromatic lens  focuses the light onto the object and the quarter-wave plate ensures that most of the light reflected from the object is transmitted through PBS-1. The polarisation optics in the reference arm ensure that the polarisation states of the photons are the same at the input of the 50:50 fibre splitter. To this end, the first half-wave plate maximises transmission through the polarisation beam-splitter PBS-2, ensuring the same polarisation as the one at the output of the object arm's PBS-1. The second half-wave plate in the reference arm is added to pre-compensate for any polarisation mismatch caused by the input fibre of the 50:50 fibre splitter. All the fibres were taped to the optic table to prevent them from moving and affecting the polarisation. The interferometer is unbalanced in terms of the chromatic dispersion due to the unequal amount of glass in the arms, mostly related to the presence of the lens in  the object arm. This imbalance is intentionally left uncorrected to show that, although the even-order dispersion responsible for the bulk of image quality loss is inherently cancelled in Quantum OCT, the odd-order dispersion can be a small but still non-negligible factor in the quality deterioration.

The 50:50 fibre splitter at the interferometer output is the beam splitter where the photons overlap and quantum interfere when the object arm length is equal to the reference arm length. The two outputs of the fibre splitter are connected to 5-kilometre-long single-mode fibre spools where the photons are delayed in a wavelength-dependent manner due to chromatic dispersion. A time-tagging device provides information on the arrival time of the photons at the Superconducting Single-Photon Detectors, with the reference point provided by the photodiode illuminated by the pulsed pump light. The time window in which the detectors wait for photons is equal to the repetition time of the pump laser, and the clock in the time-tagging device is triggered with the rising slope of the signal from the photodiode.

% measurement Td-Q-OCT
The Time-domain Q-OCT measurement is performed by sequentially translating the motorised stage at the FB3 position and recording the total number of coincidences within the acquisition time window at each position. The fibre spools are not disconnected when performing the time-domain measurements.

% measurement traditional OCT
Conventional OCT uses a pulsed laser with a centre wavelength of 1550~nm. It is fed into an input port of a second 50:50 fibre splitter whose output ports are connected to FB1 and FB2. The classical fringes are detected using an optical spectrum analyser (OSA) connected to one of the output ports of the existing 50:50 fibre splitter.

\subsection*{A-scan calculation}

The A-scan can be calculated from the joint spectrum using two equivalent approaches. The first approach was already conceptually described in \figref{fig:concept}e, where the A-scan is obtained by taking the diagonal from the two-dimensional Fourier transform of the joint spectrum. For clarity and consistency with the pre-processing algorithms - presented in the subsection "Joint spectrum pre-processing" - we propose a small modification of this approach. As shown in \figref{fig:ascan-calculation}, the joint spectrum is first rotated 45 degrees counter-clockwise so that the fringes are positioned along the X axis (\figref{fig:ascan-calculation}b), resulting in the 2D Fourier transform incorporating its elements along the X axis as well (\figref{fig:ascan-calculation}c). In such a case, an A-scan is extracted by taking half of the most central row from the module of the 2D Fourier transform (\figref{fig:ascan-calculation}d).

The second approach to calculating an A-scan gives exactly the same result, but is computationally faster as it uses a one-dimensional Fourier transformation instead of a two-dimensional Fourier transformation (which is essentially one-dimensional Fourier transformation applied to all rows of the input and then applied to its columns). In the second approach, shown in \figref{fig:ascan-calculation} (dark green arrows), the joint spectrum is also rotated 45 degrees counter-clockwise and then, all rows are averaged to create one one-dimensional spectrum. The A-scan is obtained by Fourier transforming this 1D spectrum and taking a half of the resulting transform module.

In the remainder of this article, the latter approach - as the optimum one - will be applied to obtain the A-scan from the joint spectrum, whereas the 2D Fourier transform calculated from a rotated joint spectrum will be used to show the outcomes of the joint spectrum pre-processing algorithms (found in the subsection "Joint spectrum pre-processing") and the behaviour of the structural and artefact elements (found in the section "Results").

\subsection*{Joint spectrum acquisition}

\begin{figure}[t]
    \centering
    \includegraphics[width=\columnwidth]{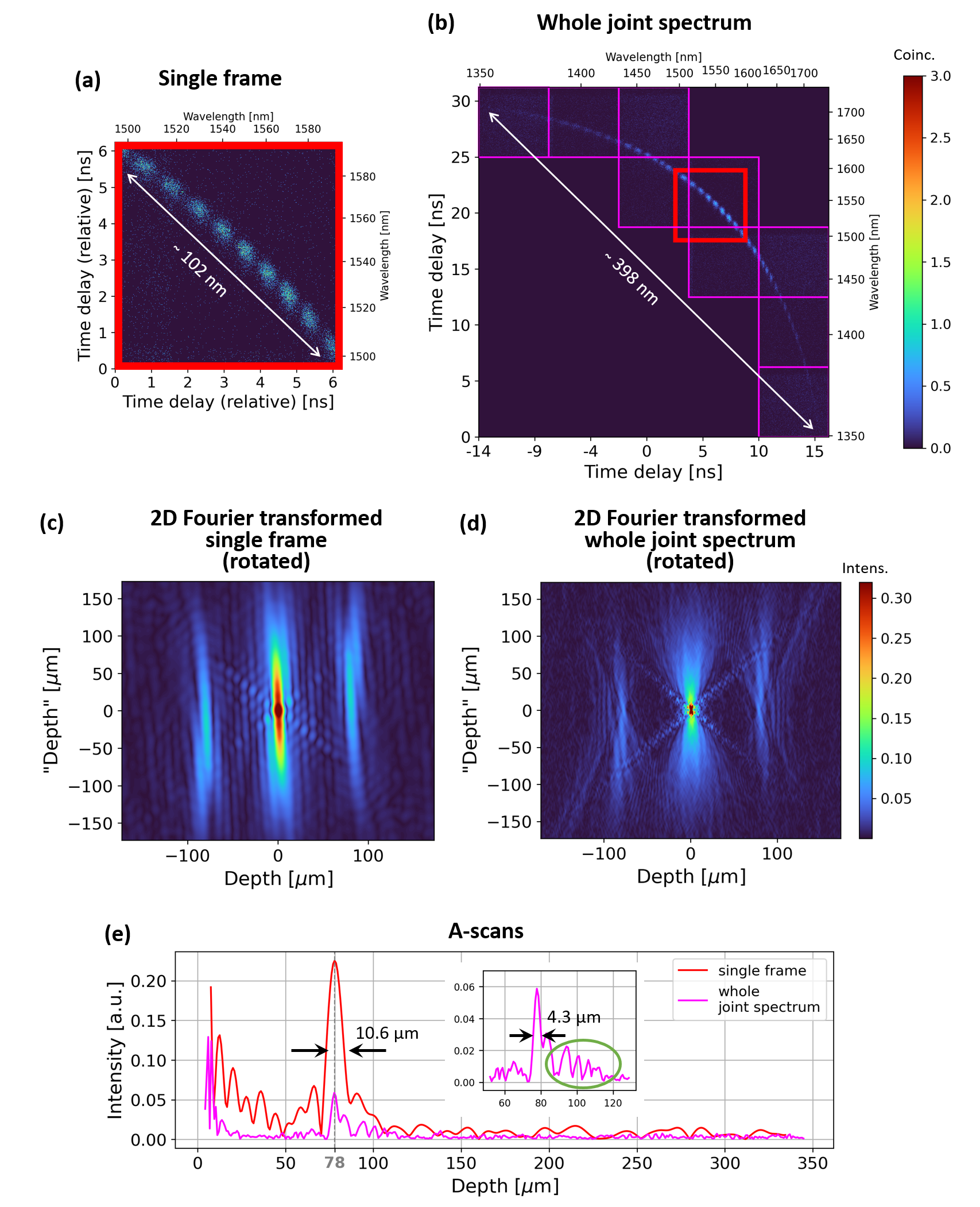}
    \caption{With a mirror as an object, (b) the whole joint spectrum covering a 398~nm wavelength range is acquired by varying the time delay in the two detection channels of the Quantum OCT setup and stitching the acquired frames. (a) A single frame, corresponding to a 102~nm wavelength range, is used for characterising the setup and comparing its performance with the conventional OCT. The anti-diagonal FWHM in both cases is 3.2~nm (see section S5). 2D Fourier transformation reveals the degrading effects of chromatic dispersion, the more so, the broader spectrally the photon pairs: (c) while the elements of the 2D Fourier transform of a single frame deviate from the expected Gaussian-like peaks (compare with \figref{fig:ascan-calculation}c), (d) the elements of the 2D Fourier transform of the whole joint spectrum are considerably distorted. (e) As expected, the A-scan resolution for the single frame (red curve), 4.3~\(\mu\)m, is worse than for the whole joint spectrum (purple curve), 10.6~\(\mu\)m. Also, uncompensated third-order dispersion is visible in the latter in the form of one-sided modulations (green circle in the inset).}
    \label{fig:whole-single}
    \vspace{-10pt}
\end{figure}

The photon pairs generated in the proposed Quantum OCT setup are spectrally broadband. As a result, the arrival times of some photons at the output of the fibre spools are larger than the coincidence time window, equal to the repetition time of the pulsed pump laser. Consequently, only a piece of the whole joint spectrum is acquired within the specified integration time during the measurement, where the specific piece to be acquired is selected by setting well-defined time delays in each detection channel using the time-tagging device. An example of such a single frame is presented in  (\figref{fig:whole-single}a), measured for a mirror as an object in the object arm and approximately 78~$\mu$m away from the zero optical path difference in the interferometer.

To measure the whole joint spectrum (\figref{fig:whole-single}b), the time delays in the detection channels are varied and the acquired frames, each containing different parts of the joint spectrum, stitched together in the post-processing (see the details of the stitching algorithm in the S1 section of the Supplementary Document). In the case of our photon pair source and fibre spools, 9 different frames are enough to show the whole joint spectrum, with each frame covering the wavelength range of around 102~nm (the wavelength calibration procedure found in section S2 of the Supplementary Document). 

Due to the large bandwidth of nearly 400~nm, the whole joint spectrum is especially affected by various chromatic dispersion effects within the experimental setup. The fibre spool dispersion leads to the bending of the joint spectrum, consequently distorting the elements in the 2D Fourier transform quite severely (\figref{fig:whole-single}c) when compared to the single-frame 2D Fourier transform (\figref{fig:whole-single}d, also compare with perfect Gaussians in \figref{fig:ascan-calculation}c). Similarly, the A-scan - unaffacted by fibre dispersion (see "Joint spectrum pre-processing" subsection for details) - is significantly distorted in the whole joint spectrum case (pink curve in \figref{fig:whole-single}e, see the characteristic one-sided oscillations in the green circle) while maintaining a fairly Gaussian-like shape in the case of a single frame (red curve in \figref{fig:whole-single}e).

The single-frame joint spectrum best balances the peak quality and the acquisition time as the detrimental impact of different chromatic dispersion sources is minimised in it and a single frame is acquired 9 times faster. Also, because the wavelength range covered by a single frame matches best the wavelength range of the laser used to perform conventional OCT, the single frame is considered and used for analysis and fair comparisons of Quantum OCT with the conventional OCT.

However, to have a full picture of the capabilities of the used Quantum OCT setup, we also present the key performance metrics for when the whole joint spectrum is used, including the analysis of the axial resolution and the imaging range (found in the subsection "Axial resolution and imaging range" of the "Results" section). Also, the whole joint spectrum is used to visualise and develop pre-processing algorithms, described in the following subsection "Joint spectrum pre-processing".

\subsection*{Joint spectrum pre-processing}

An experimental joint spectrum, whether a single frame or a whole joint spectrum, is pre-processed using two algorithms: fibre dispersion compensation algorithm and pump-related dispersion compensation algorithm.

\begin{figure}[htb]
    \centering
    \includegraphics[width=\columnwidth]{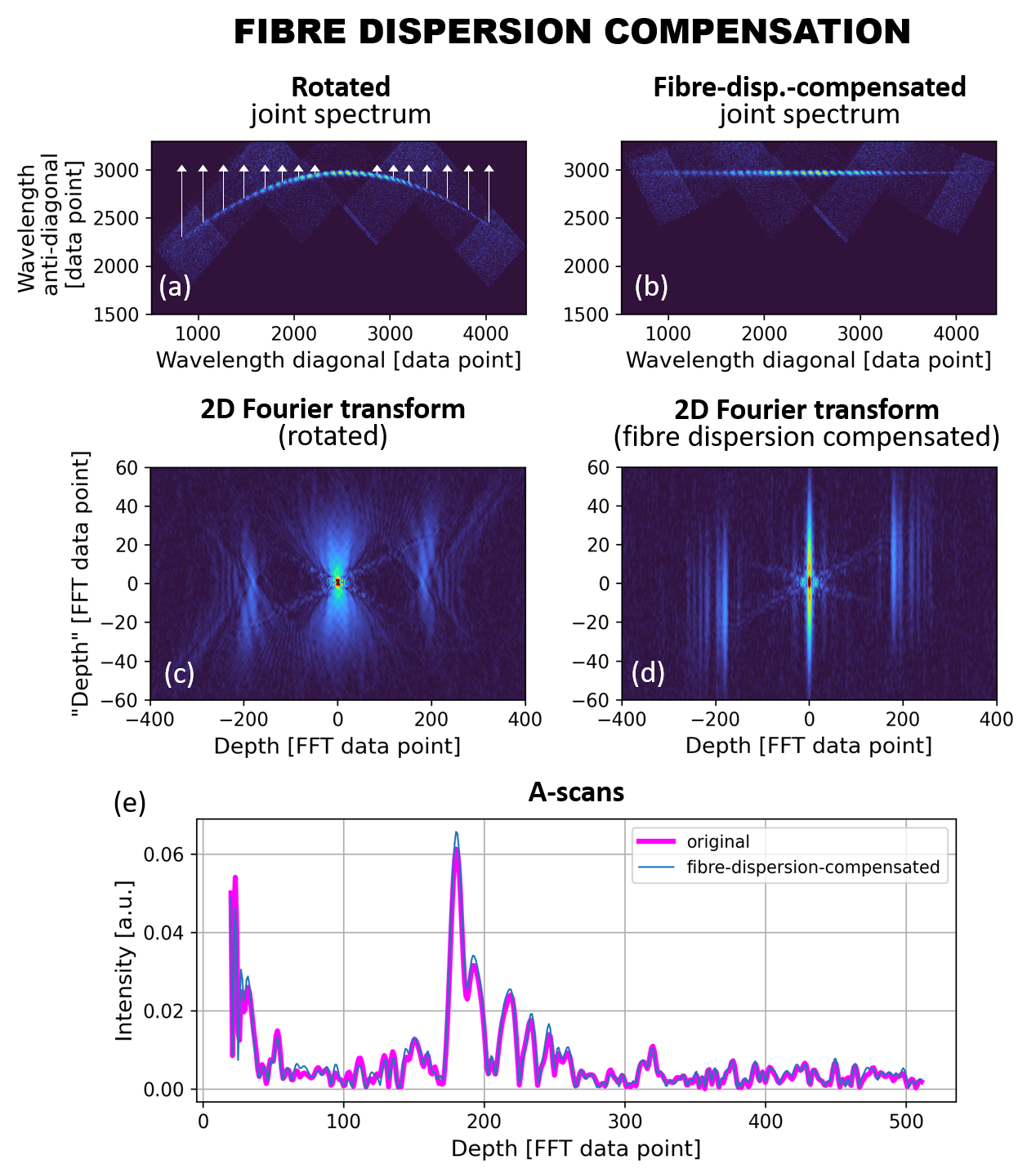}
    \caption{ Applying a fibre-dispersion-compensating vector - conceptually represented by white arrows - (a) to a target joint spectrum whose (c) 2D Fourier transform is of a poor quality results in (b) a straightened joint spectrum whose (d) 2D Fourier transform shows high-quality sharp elements. Because the fibre dispersion compensation does not change the average statistics of the joint spectrum, (e) the original A-scan, plotted in pink, and the fibre-dispersion-compensated A-scan, plotted in blue, are the same.}
    \label{fig:fibre-dispersion-compensation}
%    \vspace{-10pt}
\end{figure}

\begin{figure}[htb]
    \centering
    \includegraphics[width=\columnwidth]{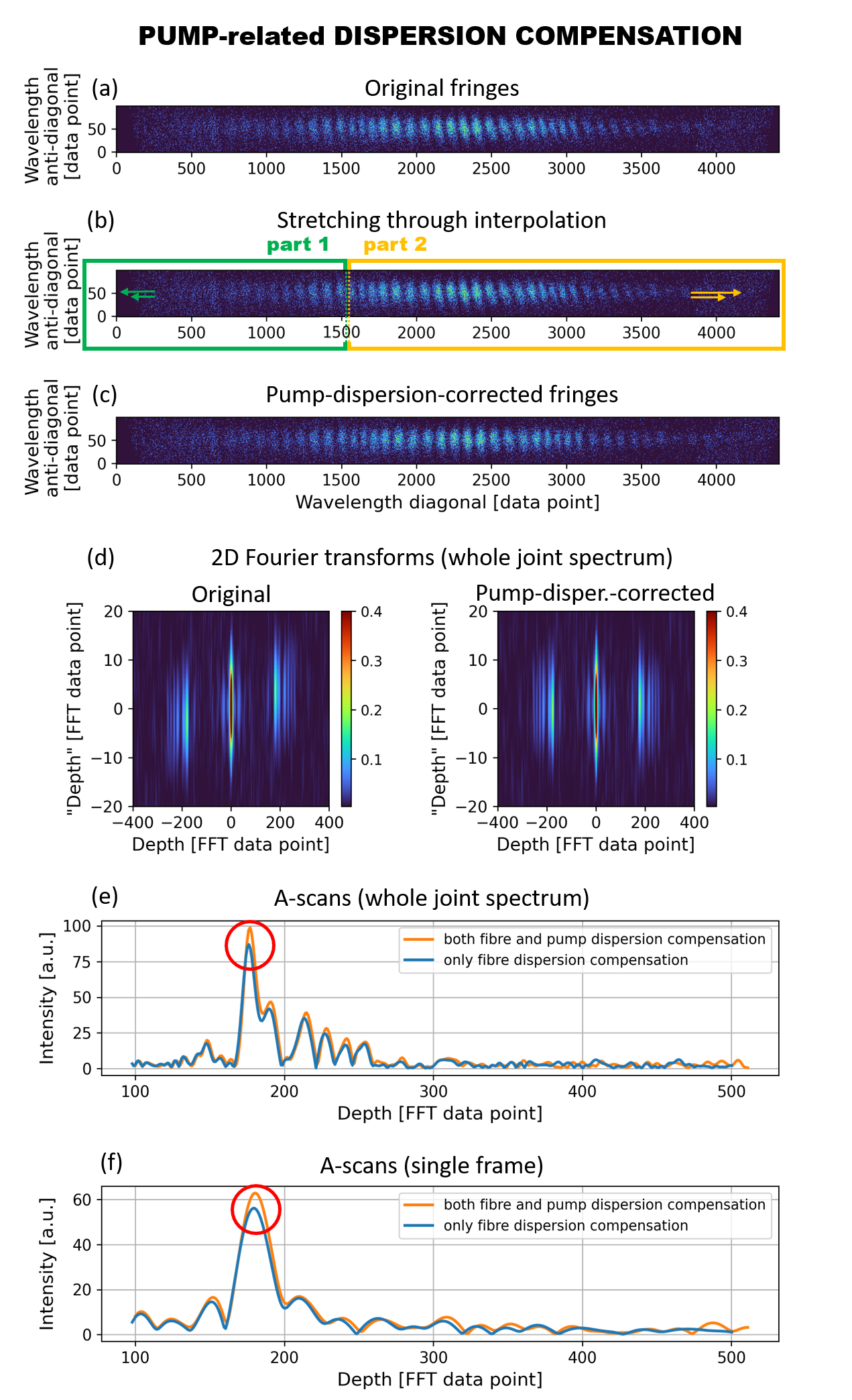}
    \caption{As best seen on single-frequency fringes generated by a mirror as an object, (a) the fringes frequency varies diagonally, introducing dispersion-like effects. (b) Fringes frequency is equalised by dividing them into two parts and stretching each part by means of interpolation. As a result, (c) the fringes become uniform, producing (d-right) symmetrical 2D Fourier transform as compared to (d-left) 2D Fourier transform of the original fringes, and (e) an A-scan with an improved height (orange curve) as compared to the A-scan obtained using the original fringes (blue curve). (f) The height is also improved in the case of an A-scan obtained from a single frame. }
    \label{fig:pump-dispersion-compensation}
%    \vspace{-20pt}
\end{figure}

\textbf{Fibre dispersion compensation} reverses the effect of the chromatic dispersion of the 5-kilometre-long fibre spools used in the detection system by shifting the data points to their correct positions, leading to the straightening of the joint spectrum.
In theory, the shifting should be carried out independently in the vertical direction and in the horizontal direction of the joint spectral plane, as each represents a different detection channel. In practice, since both channels
%are identical in  terms of the fibre length and therefore the accumulated dispersion, 
have similar fibre lengths and therefore the accumulated dispersion
one can correct simultaneously for both by rotating the joint spectrum by 45 degrees counter-clockwise (\figref{fig:fibre-dispersion-compensation}a) and rolling the columns by a correct number of places (\figref{fig:fibre-dispersion-compensation}b). The result can then be rotated by 45 degrees clockwise so the joint spectrum is presented in the original wavelength axes corresponding to the wavelengths of photons detected in each channel. Details of this algorithm are found in section S3 of the Supplementary Document.

Fibre dispersion compensation restores high quality of the elements in the 2D Fourier transform (compare \figref{fig:fibre-dispersion-compensation}c and d). The A-scan corresponding to the original joint spectrum, purple curve in \figref{fig:fibre-dispersion-compensation}e, is almost identical to the A-scan obtained from the fibre-dispersion-corrected joint spectrum, blue curve in \figref{fig:fibre-dispersion-compensation}e. It is an expected result as rolling the columns does not affect the outcome of a row-wise averaging used for calculating the A-scan in both cases. Although fibre dispersion compensation does not introduce any improvement in the A-scan, it is a crucial step nonetheless as it enables pump-related dispersion compensation. Apart from that, it allows an interpretation of the 2D Fourier transform in terms of the behaviour of the artefacts in the case of more complicated objects such as a piece of glass.

\textbf{Pump-related dispersion compensation} addresses the peak quality degradation introduced by a spectrally broadband pump, where a different pump wavelength produces a different diagonal in the acquired joint spectrum. As observed in a joint spectrum obtained with a mirror as an object (rotated and fibre-dispersion compensated in \figref{fig:pump-dispersion-compensation}a), fringes in each row (each diagonal in the unrotated version) have a slightly different frequency. During row-wise averaging - a step required to obtain an A-scan, the wash-out of fringes occurs, reducing their visibility at the sides and consequently producing a peak with a reduced height and broadened width in the Fourier transform (see \figref{fig:pump-dispersion-compensation}e and f for the former effect). In this way, although each row (each diagonal in the joint spectrum's unrotated version) individually represents an even-order-dispersion-free signal - so a signal producing no peak broadening or height reduction related to the second-order dispersion, the set of diagonals will produce these outcomes, weakening the dispersion cancelling effect of Quantum OCT. This effect is the bigger, the broader spectrally the pump is and the broader spectrally the photon pairs are \cite{okano2013dispersion}.

Impossible to correct in Time-domain Q-OCT, the varying frequency of the rows (diagonals in the unrotated version) is easily adjusted in Fourier-domain Q-OCT. Here, we do it by splitting the area of the joint spectrum containing the fringes into two parts and stretching each part individually  using interpolation (\figref{fig:pump-dispersion-compensation}b), producing pump-dispersion-corrected fringe pattern, presented in \figref{fig:pump-dispersion-compensation}c (further details found in section S4 of the Supplementary Document). As shown in \figref{fig:pump-dispersion-compensation}d, pump dispersion compensation further improves the quality of the 2D Fourier transform, restoring its symmetricity with regards to the central zero-depth peak.

The pump bandwidth - although considered broad using quantum optics standards - was not broad enough to substantially affect the peak quality and led only to a slight improvement of the peak height in the case of both the whole joint spectrum (\figref{fig:pump-dispersion-compensation}e) and a single frame (\figref{fig:pump-dispersion-compensation}f). This finding validates the experimental results of \cite{graciano2019interference} where a near-perfect dispersion cancellation was obtained using a Time-domain Q-OCT with a similar photon pair source.

%As shown in more detail in \figref{fig:fall-off}a-d in the following section, the peak height gets the more improved, the deeper it is in the A-scan (the denser the fringes in the joint spectrum), leading to enhanced imaging range for both the whole joint spectrum and the single frame. Since the imaging range is very limited by default in Fourier-domain setups due to its relationship with the detection's spectral resolution \cite{hu2007analytical}, both fibre and pump dispersion compensation is applied to the acquired data.

\begin{figure*}[htb]
    \centering
    \includegraphics[width=\linewidth]{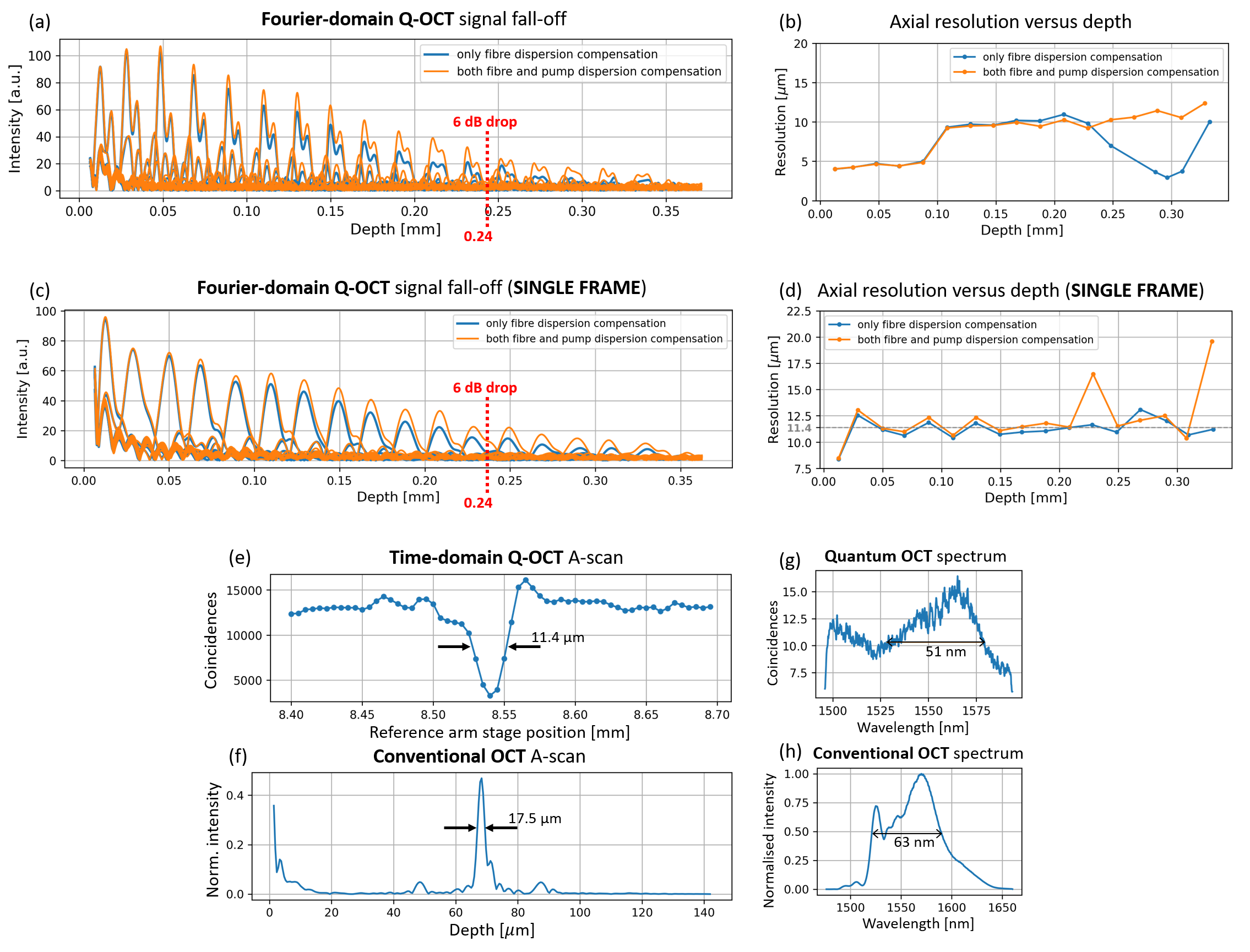}
    \caption{(a) Signal fall-off calculated using the whole joint spectrum, and (c) the signal fall-off calculated using a single frame indicate that the effective imaging range improves from 0.19~mm to 0.24~mm when the pump dispersion compensation is performed. While (b) the axial resolution worsens with depth for the whole-joint-spectrum case due to the unbalanced third-order dispersion, (d) the axial resolution remains stable for the single-frame case being around 11.4~$\mu$m. (e) As expected, the same axial resolution is obtained when the signal is acquired in the time domain, but (f) it is worse, 17.5~$\mu$m, in the conventional OCT, with (h) conventional OCT spectrum having a spectral bandwidth of 63~nm, similar to (g) the diagonal width of the joint spectrum, 51~nm.}
    \label{fig:fall-off}
    %\vspace{-15pt}
\end{figure*}

\section*{Results}

This section starts with an analysis of two parameters of the proposed Q-OCT setup: the axial resolution and the imaging range, being the key characteristics representing the general imaging performance of a Fourier-domain OCT system. This analysis is provided for two cases: the whole joint spectrum and a single frame. In each case, a comparison is presented of the results obtained when only fibre dispersion compensation is performed and when both fibre dispersion and pump-related dispersion compensation are performed, helping to better assess the extent of the weakening of the dispersion cancellation effect due to the broadband pump. A-scans showing the axial resolution for Time-domain Q-OCT and for conventional OCT, which uses a similar wavelength range as the single-frame case, are provided for further comparisons.

The actual imaging performance of the proposed Fourier-domain Q-OCT setup is tested by imaging 3 objects: a thin piece of glass, a thick piece of plastic and two thin pieces of glass. Again, a single-frame joint spectrum is acquired during these measurements so that a fair comparison with conventional OCT can be made. A joint spectrum in each case is both fibre- and pump-dispersion compensated and presented in its original diagonal form, with the 2D Fourier transform shown in the rotated version for the ease of interpretation. Only Fourier-domain Q-OCT A-scans obtained from both fibre- and pump-dispersion-compensated joint spectra are presented (A-scans obtained from joint spectra where only fibre-dispersion compensation was applied can be found in section S6 of the Supplementary Document, together with the dispersion-uncompensated joint spectra and their corresponding 2D Fourier transforms).

%but also because it best balances the acquisition time and the peak quality. As already seen in \figref{fig:whole-single}, on one hand, the full joint spectrum - consisting of 7 frames - takes 7 times longer to acquire compared to the single frame. On the other hand, the full joint spectrum gives a better axial resolution, but is affected by the unbalanced third-order dispersion which leads to the peak distortions in the form of one-sided modulations, absent in the single-frame case.

\subsection*{Axial resolution and imaging range}

The assessment of the axial resolution and imaging range of the Fourier-domain Q-OCT setup is performed using joint spectra acquired for 17 different lengths of the reference arm with the mirror as an object in the object arm. The joint spectra are processed using the approach presented in \figref{fig:ascan-calculation} (dark green arrows) into A-scans  which when plotted together in \figref{fig:fall-off}a~and~c show the signal fall-off, i.e. depth-dependent sensitivity decay due to the detection's finite spectral resolution \cite{hu2007analytical}.
While \figref{fig:fall-off}a shows the signal fall-off for when the whole joint spectrum is used, \figref{fig:fall-off}c shows it for the single frame, with the blue curve peaks being the result of only the fibre dispersion compensation and the orange curve peaks - the result of both the fibre and pump dispersion compensation.

One can see that the effective imaging range, defined as the depth at which a 6-dB drop in the peak height occurs, becomes  longer when the pump dispersion compensation is carried out, improving this parameter from 0.19~mm to 0.24~mm for both the whole joint spectrum and the single frame. The imaging range is the same for these two cases because it does not depend on the spectral bandwidth of the joint spectrum, but rather on the detection's spectral resolution - the same in both cases.

The axial resolution - calculated as the peak's full width at half maximum (FWHM) - in the whole-joint-spectrum case (\figref{fig:fall-off}b) is below 5~$\mu$m for shallow depths and then degrades to over 10~$\mu$m at 0.1~mm due to the influence of the unbalanced third order dispersion in the interferometer. More specifically, the one-sided oscillations appearing due to the  third-order dispersion become higher than the peak's half-maximum and consequently, start to contribute to the FWHM. The improvement of the axial resolution after 0.2~mm depth in the case where only fibre dispersion compensation was performed is due to the more favourable shape of the distortions caused by the third-order dispersion.
On the other hand, when only a single frame is used (\figref{fig:fall-off}d), the axial resolution remains at more or less stable level of 11.4~$\mu$m.
The theoretical values of the axial resolution, 10.4~$\mu$m for the single-frame case and 2.9~$\mu$m for the whole-joint-spectrum case (details on how they were calculated are found in section S5 of the Supplementary Document), are lower than the experimental ones by 10\% and 72\%, respectively. This is due to the unbalanced third-order dispersion in the interferometer, which - apart from inducing one-sided oscillations - broadens the peak, the effect the more prominent, the broader the spectral bandwidth of the light.

The axial resolution in Fourier-domain Q-OCT is compared to the axial resolutions obtained using the Time-domain Q-OCT approach and conventional OCT with a similar spectral bandwidth. The Time-domain Q-OCT signal (\figref{fig:fall-off}e) was acquired by changing the length of the reference arm with a step size equal to 5~$\mu$m, and recording the total number of coincidences. The resulting dip's FWHM is 11.4~$\mu$m, which, as expected, matches the peaks' average FWHM in the Fourier-domain case.

The conventional OCT spectrum was acquired, nonlinearity-corrected using the algorithms from Ref.~\cite{wang2008spectral} and Fourier transformed. The resulting A-scan, presented in \figref{fig:fall-off}f, shows a peak whose FWHM is approximately 17.5~$\mu$m. As already shown both theoretically and experimentally in previous work on this subject, the conventional OCT axial resolution is worse than that obtained with Q-OCT for the similar spectral bandwidth. In our case, the increase of the axial resolution is not exactly two times because of slight differences between the spectral bandwidth of the laser and the photon pairs' bandwidth. Whereas the FWHM of the laser spectrum  in conventional OCT is approx.~63~nm  and the central wavelength approx. 1580~nm (\figref{fig:fall-off}h), the diagonal FWHM of the single-frame joint spectrum  is approx.~51~nm and the central wavelength approx.~1553~nm (\figref{fig:fall-off}g), producing 7.6~THz and 6.3~THz frequency bandwidths, respectively. This difference of approx.~20$\%$ results in the conventional OCT axial resolution being slightly better than twice the value of the quantum OCT resolution.

One single-frame joint spectrum was acquired in 2 seconds, while each point in the time-domain signal was acquired for 1 second, resulting in a total acquisition time of around 2 minutes. The conventional OCT spectrum was measured using the OSA in under a second which is considered rather long when compared to the acquisition times achieved by standard OCT machines. Please see Tab.~\ref{tab:time} for the comparison.

\begin{figure}[htb]
    \centering
    \includegraphics[width=0.95\linewidth]{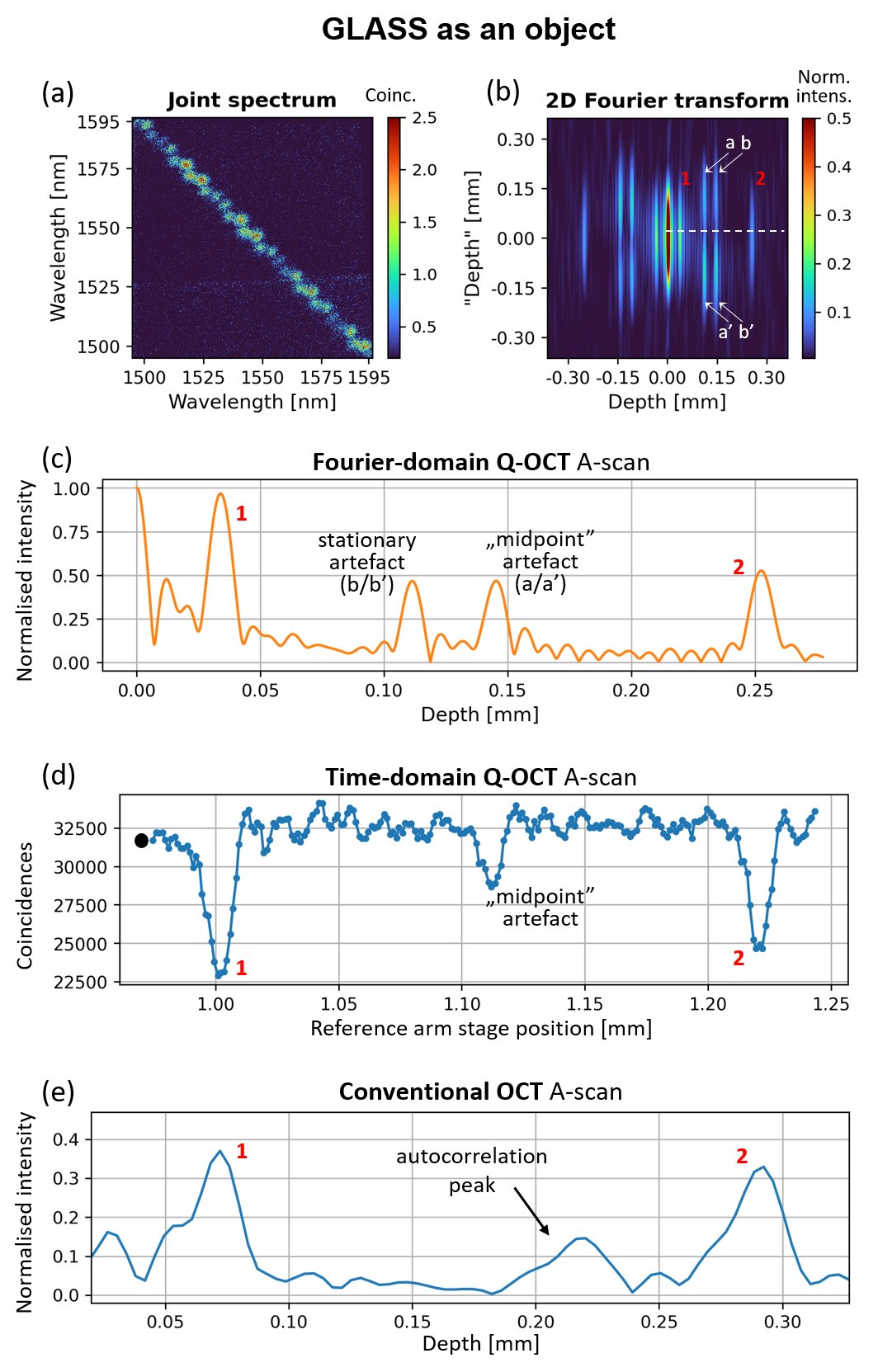}
    \caption{Measurement of the glass sample. (a) Fibre- and pump-dispersion-corrected joint spectrum produces (b) 2D Fourier transform with sharp, undistorted peaks: two corresponding to the structure of the imaged glass (1,2) and four being the artefacts (a, a' and b, b'). (c) Pump-dispersion compensation produces Fourier-domain Q-OCT A-scan (orange curve) with higher peaks than when only fibre-dispersion compensation is applied (blue curve). Both show two clear structural peaks and two artefact peaks b/b' and a/a'. (d) The time-domain A-scan shows structural dips as well as an artefact, and (e) the conventional OCT A-scan shows the structure of the object with a lower resolution and with an autocorrelation peak.  }
    \label{fig:linearisation-glass}
   %\vspace{-15pt}
\end{figure}

\subsection*{Single-layer glass}

Using layered structures as objects gives more insight into the appearance of artefacts, i.e. peaks not related to the structure of the imaged object. Artefacts are inseparable with Quantum OCT, originating from the interfering photons reflected from the front and back surfaces of object layers \cite{kolenderska2020quantum,kolenderska2020fourier}.

First, we use a 100-$\mu$m-thick glass as an object and acquire a single-frame joint spectrum for 3 minutes at the reference arm length being 50~$\mu$m shorter than the object arm length (the front surface of the glass serving  as the object arm's end). The reference stage position at which the joint spectrum is measured is marked with a black dot in the time-domain signal in \figref{fig:linearisation-glass}d. The acquired joint spectrum is fibre and pump dispersion compensated and plotted in its diagonal form in \figref{fig:linearisation-glass}a.

In the 2D Fourier transform (\figref{fig:linearisation-glass}b), one can discern the zero-depth peak in the centre and two structural peaks (marked with red 1 and 2) located in the central row which corresponds to the joint spectrum diagonal. There are four additional peaks situated symmetrically around the central row, \textit{a} and its twin \textit{a'} at the midpoint between the structural peaks and \textit{b} and its twin \textit{b'} at the distance from the zero-depth peak equal to half the glass thickness. The same set of peaks appears in the other half of the 2D Fourier transform just as in the case of the mirror (\figref{fig:pump-dispersion-compensation}d).

%\textcolor{red}{Cris (I would include some explanation why the artifacts behave as they do, if it is to obvious for your community, disregard it): 
%In the case of two reflecting interfaces at depths \(z_{1}\) and \(z_{2}\), the detected two-photon interference signal contains not only the structural contributions from each interface separately but also cross-terms arising when one photon is reflected from \(z_{1}\) and the other from \(z_{2}\). These cross-terms correspond to products of waves with delays \(z_{1}\) and \(z_{2}\), and give rise to oscillatory factors proportional to \(\cos\!\big(k(z_{1}-z_{2})\big)\) and \(\cos\!\big(k(z_{1}+z_{2})\big)\). After Fourier transformation, these terms manifest as artefacts at positions corresponding to half the separation \(|z_{2}-z_{1}|/2\) and the midpoint \((z_{1}+z_{2})/2\) of the two interfaces, respectively. The former is independent of the absolute reference arm position and therefore appears as a stationary artefact, while the latter depends on the absolute delays and shifts with the reference arm, leading to the so-called ``midpoint'' artefact.}

Naturally, the Fourier-domain Q-OCT A-scan (\figref{fig:linearisation-glass}c) incorporates artefacts as well: a "midpoint" one associated with \textit{a}/\textit{a'}, and a stationary one associated with \textit{b}/\textit{b'}. This is due to the fact that \textit{a} and \textit{a'} (as well as \textit{b} and \textit{b'}) overlap the horizontal zero-depth axis (dashed white line in \figref{fig:linearisation-glass}b). This confirms our simulation result \cite{kolenderska2020fourier} which predicted two artefacts per interface pair: the "midpoint" one - also found in the Time-domain Q-OCT A-scan - appearing halfway between the layer interfaces, and the stationary one - appearing only in Fourier-domain Q-OCT - found at a position equal to half the layer thickness.

For reference, a time-domain signal (\figref{fig:linearisation-glass}d) was acquired  by counting coincidence events for a range of reference stage positions (2.5~$\mu$m step, 10 seconds integration per data point, resulting in over 30-minute total acquisition time). Such a Time-domain Q-OCT A-scan incorporates dips corresponding to the glass structure (1, 2 in \figref{fig:linearisation-glass}d). As expected, there is an additional, artefact dip at the midpoint between the structural dips.

The conventional OCT A-scan, in \figref{fig:linearisation-glass}e, shows both peaks as well as an autocorrelation peak, which is a peak appearing due to the interference of light reflected from the objects' surfaces. As expected, the structural peaks' width is larger than that of the peaks and dips in the Q-OCT signals. The conventional OCT spectrum was acquired using the OSA, again, in less than a second (performance comparison in Tab.~\ref{tab:time}).

\subsection*{Single-layer plastic}

\begin{figure}[ht!]
    \centering
    \includegraphics[width=0.95\linewidth]{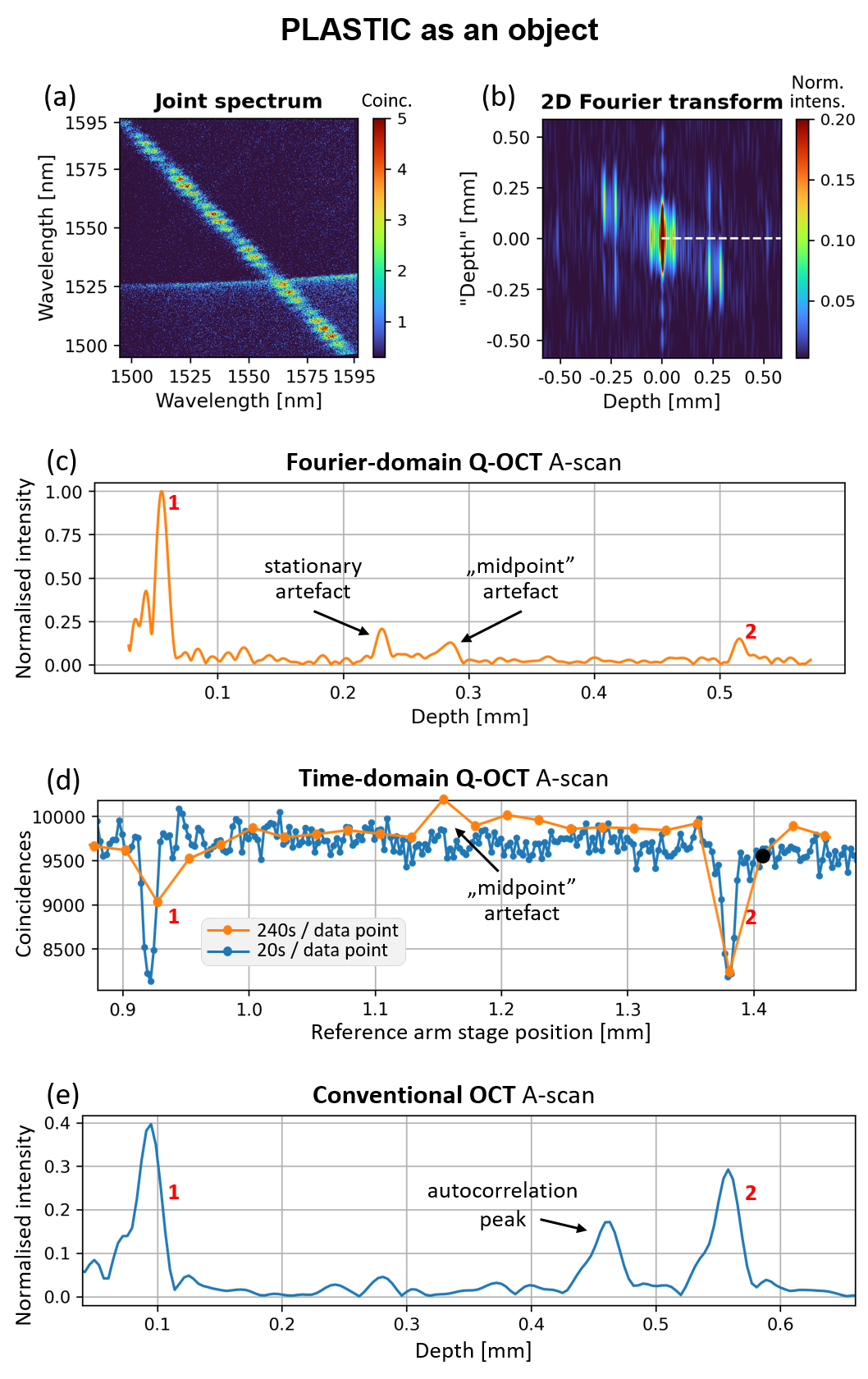}
    \caption{Measurement of the plastic sample. (a) Fibre- and pump-dispersion-corrected joint spectrum is (b) Fourier transformed. (c) Pump-dispersion compensation produces higher peaks in the A-scan (orange curve) as compared to the situation when only fibre-dispersion is applied (blue curve). (d) The "midpoint" artefact is seen in the long-acquisition time-domain A-scan (orange), but is gone for short-acquisition time-domain A-scan (blue) due to noise. (e) Conventional OCT A-scan shows the structure of the object with a lower resolution and with an autocorrelation peak. }
    \label{fig:plastic}
   %\vspace{-20pt}
\end{figure}

A piece of plastic, which is 260~$\mu$m thick, was used as an object and the joint spectrum was acquired for 4 minutes at the reference arm length being 45~$\mu$m bigger than the object arm length (the back-surface of the plastic serving in this case as the object arm's end). The reference arm stage position at which the joint spectrum was measured is marked with a black dot in the time-domain signal in \figref{fig:plastic}d. The acquired joint spectrum is fibre and pump dispersion compensated (\figref{fig:plastic}a) and 2D Fourier transformed (\figref{fig:plastic}b). One observes a horizontal element in the joint spectrum, related to the pump photons leaking into the interferometer, most probably caused by a setup misalignment in these measurements. As the leaked pump photons are not considerably dispersed by a fibre spool in one channel, they map to a nearly fixed arrival time when coinciding with an accidental count in the other channel. This produces a line-like accumulation of coincidences. While here, one sees a vertical accumulation, a horizontal one is also present, as it can be deduced from a closer inspection of the whole joint spectrum in \figref{fig:fibre-dispersion-compensation}a. 

The Fourier-domain Q-OCT A-scan in \figref{fig:plastic}c shows two peaks corresponding to the front and back surface of the plastic (marked with 1 and 2). Due to the small imaging range, the back-surface peak height is very small. There are two artefacts, just as in the case of the glass, the "midpoint" one located halfway between the structural peaks and stationary one located at the distance from zero-depth peak equal to the plastic thickness. The height of the artefact peaks is much smaller with respect to the structural peaks than the artefacts for the glass (\figref{fig:linearisation-glass}c), because the plastic is thicker than the glass. This confirms the theoretical findings in Ref. \cite{kolenderska2021artefact} where it was noted that artefact suppression is the better, the bigger the anti-diagonal width of the joint spectrum and the thicker the object. Since the joint spectrum's anti-diagonal width is the same for all the imaged objects, the difference in the reduction of the artefact height comes from the object thickness.

Two Time-domain Q-OCT A-scans were acquired: a coarse one where the reference arm stage position was varied by 50~$\mu$m and the integration time was 4 minutes per data point  (\figref{fig:plastic}d, orange curve), and a fine one with a much smaller step, 5~$\mu$m, and much shorter integration time per data point, 20 seconds (\figref{fig:plastic}d, blue curve). While the coarse Time-domain Q-OCT A-scan undersampled the structural dips, it was able to capture the "midpoint" artefact which in this case is a peak, not a dip. On the other hand, in the fine Time-domain Q-OCT A-scan, the structural dips are clearly visible and no artefact is observed as it is buried in the noise due to the insufficient integration time. This result shows that Fourier-domain Quantum OCT and time-domain Quantum OCT are equivalent in terms of the information they provide as long as the acquisition time of the former method is equal to the integration time of a single data point in the latter method (see an extended analysis in section S7 of the Supplementary Document).
With a total of 258 points, the acquisition time of the fine scan was over 1.5 hours, while the coarse one - comprising 24 data points - took over 1.6 hours to complete.

%\textcolor{red}{Cris: why one the structural and midpoint artifact are diagonal, whereas the stationary is static? The stationary art. only depends on the spacing between the two interfaces, not their absolute position. It is then vertical for it is located at teh same depth for all reference arm settings. right?}

The conventional OCT A-scan, in \figref{fig:plastic}f, shows both peaks as well as an autocorrelation peak with the widths larger than those of the peaks and dips in Q-OCT signals. Again, the conventional OCT spectrum was acquired in less than a second.

\begin{figure}[!htb]
    %\centering
    \includegraphics[width=213pt]{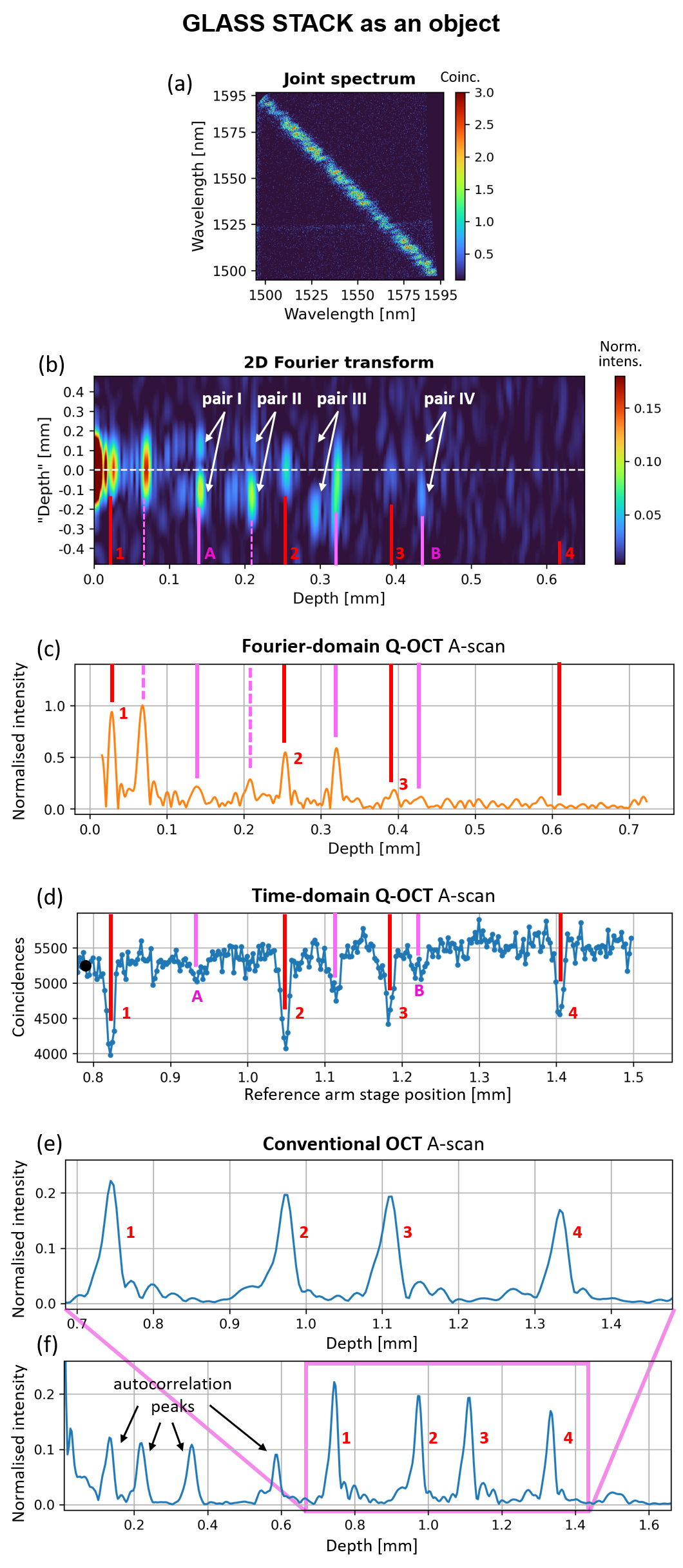}
    \caption{Measurement of a stack of thin glasses. (a) Fibre- and pump-dispersion-compensated joint spectrum and (b) the positive half of its 2D Fourier transform showing artefact pairs I, II, III and IV whose twin visibility decreases with depth. (c) The Fourier-domain Q-OCT A-scans show three, out of the expected four, structural peaks - marked 1, 2 and 3, and five artefacts. (d) In the time-domain signal, all 4 structural elements are observed as dips, with one clear and two less clear artefact dips. (e) Conventional OCT A-scan shows the structure of the object with a lower resolution and with autocorrelation peaks cluttering the front of (f) the A-scan. Red lines mark the structural elements and the purple lines mark the artefact elements: dashed ones pointing the Fourier-domain Q-OCT specific stationary artefacts, and the solid ones pointing to the "midpoint" artefacts.  }
    \label{fig:2glass}
    \vspace{-15pt}
\end{figure}

\subsection*{Glass stack}

Finally, a stack of two 100-$\mu$m-thick glasses was put as an object in the object arm and the joint spectrum was acquired for 4 minutes at the reference arm length being 35~$\mu$m shorter than the object arm length (the front surface of the stack marking the object arm end). Again, the reference stage position at which the joint spectrum was measured is marked with a black dot in the time-domain signal in \figref{fig:2glass}d).

The acquired joint spectrum was fibre and pump dipsersion corrected (\figref{fig:2glass}a) and then 2D Fourier transformed (\figref{fig:2glass}b). Because the object comprises 4 interfaces (red vertical lines) - front and back surfaces of two pieces of glass - the 2D Fourier transform, whose only the positive half is shown for clarity, is cluttered with artefact elements (purple vertical lines).
The Fourier-domain A-scan, \figref{fig:2glass}c, incorporates artefacts as well. Also, due to the limited imaging range, the fourth peak - corresponding to the back surface of the second piece of glass in the stack - is missing.

The time-domain A-scan, \figref{fig:2glass}d, with 30 seconds integration time per all 300 data points each, was acquired in over 2.5 hours. In it, all four structural elements are clearly visible in the form of dips. Because the distance between interface 2 and interface 3 is small, the artefact corresponding to that pair is clearly seen. A comparison with the 2D Fourier transform and Fourier-domain Q-OCT A-scan lets one identify two more artefact dips despite their low visibility due to the short integration time: artefact A at the midpoint between interface 1 and 2, and artefact B between interface 2 and 4.

A further analysis of the elements in the 2D Fourier transform leads to a conclusion that the artefact pairs, marked as pair I-IV in \figref{fig:2glass}b, become less symmetrical the deeper they are, with the twin disappearing completely for pairs III and IV. Such asymetricity is also observed in the 2D Fourier transform representing the plastic (\figref{fig:plastic}b) and is most probably connected to the fibre dispersion whose influence increases with the depth.

In the conventional OCT A-scan, \figref{fig:2glass}e, again acquired in less than 1 second, one observes all four peaks. In this case, the reference stage position had to be chosen in a way such that the structural peaks do not overlap the autocorrelation peaks which clutter the front of the A-scan, see \figref{fig:2glass}f.

\section*{Summary and discussion}

We have provided an experimental demonstration of Fourier-domain Q-OCT, a modality theoretically proposed and analysed by us in 2020 \cite{kolenderska2020fourier}. The introduction of 5-kilometre-long fibre spools in the two detection Q-OCT channels allowed us to measure a two-dimensional joint spectrum of the interfering photon pairs for a variety of objects: a mirror, a thin piece of glass, a thick piece of plastic and a stack of two glasses.
Two post-processing algorithms were developed, compensating the degrading effects of chromatic dispersion originating from the fibre spools and the broadband pump. Their application allowed to obtain high-quality 2D Fourier transforms enabling identification of the artefacts, especially useful for complex objects (here, represented by a stack of glasses, \figref{fig:2glass}b-d).

As expected, the Fourier-domain Q-OCT A-scans showed the axial resolution much better than that obtained with the conventional OCT employing light of a similar bandwidth (\figref{fig:fall-off}). The axial resolution was not degraded by the unbalanced even-order dispersion introduced by the excess amount of glass present in the object arm (\figref{fig:setup}), but it was nonetheless affected by the unbalanced third-order dispersion, especially in the case of the spectrally broadband whole joint spectrum acquisition. This result provides a reality check for the dispersion-cancelling feature of Quantum OCT. The effects of too huge dispersion imbalance in the interferometer - whether originating from the nonsymmetricity in optical components or caused by the object itself - will not be completely cancelled in Q-OCT, because the contribution of the odd-order dispersion will become more and more significant. Although odd-order dispersion can be neglected for sufficiently narrowband light (as it was proven the case in the single-frame joint spectrum acquisition), it needs to be accounted for for broadband light. This is especially important as the broader spectrally the light, the higher the axial resolution.

\begin{table*}[ht]
    \centering
    \begin{tabular}{|c|c|c|c|c|}
    \hline
         & Mirror & Thin glass & Thick plastic & Glass stack \\ \hline
        \textbf{Fourier-domain Q-OCT} & \textbf{2 s} & \textbf{3 min} & \textbf{4 min}  & \textbf{4 min} \\ \hline
        Time-domain Q-OCT & 1 min (1 s)  & 30 min (10 s) & 1.5 hrs (20 s)  & 2.5 hrs (30 s) \\ \hline
        Conventional OCT & < 1 s  & < 1 s & < 1 s & < 1 s  \\
    \hline
    \end{tabular}
    \caption{Total acquisition times for different OCT modalities presented in this article. Integration time per data point in the Time-domain Q-OCT case is given in the brackets.}
    \label{tab:time}
    \vspace{-15pt}
\end{table*}

%Verification, problem and potential solution
Our measurements validated the simulations from both Ref.~\cite{kolenderska2020fourier} and Ref.~\cite{kolenderska2021artefact}: the 2D Fourier transforms of the joint spectra gave the expected insight into the limitations of the Q-OCT technique. The axial resolution depended on the diagonal width of the joint spectrum whereas its anti-diagonal width affected the artefact removal capabilities. As observed in the A-scans for the thin glass, \figref{fig:linearisation-glass}e, and a thicker plastic,\figref{fig:plastic}c, the anti-diagonal width allows a substantial reduction of the artefact peaks for the thicker plastic, but is not enough to affect the artefacts for the thinner glass. When corresponding 2D Fourier transforms are inspected, one sees that the artefact elements in there are very close to each other in the thin glass case, overlapping each other and consequently, producing non-zero values at the main diagonal. A remedy in this situation would be to reduce the anti-diagonal width of the artefact elements by employing a photon pair source whose antidiagonal joint-spectral width is larger.

%FD better handles artefacts and allows to use ML for artefacts
It was also confirmed that Fourier-domain Q-OCT produces two artefact peaks per interface pair, one at the midpoint between the interfaces ("midpoint" artefact) and one at the position equal to half the distance between the interfaces (stationary artefact), as compared to one artefact (the "midpoint" one) in Time-domain Q-OCT.
A further comparison with the Time-domain Q-OCT signal showed that the artefacts are equally suppressed for both modalities. This means that if one observes no artefacts in the time-domain signal, the Fourier-domain one will be artefact-free as well, and vice versa. The advantage of the Fourier-domain approach in this respect lies in the potential presented by the joint spectrum which gives unique access to the behaviour of the artefacts. As shown in \cite{maliszewski2021artefact, maliszewski2022removing}, when the joint spectrum is processed in a proper way and used as an input to a neural network, the artefacts could be easily identified and completely removed, a feature impossible with the time-domain signal.
Also, Fourier-domain Q-OCT achieves better sensitivity than its time-domain counterpart. As shown on the example of the plastic, \figref{fig:plastic}, Fourier-domain Q-OCT was able to capture a nearly-suppressed artefact after a 4-minute acquisition, whereas the time-domain signal - acquired for 1.5 hrs - could not as this artefact remained buried in the noise. Although in this particular case the artefact absence is a very welcomed outcome, one could imagine a scenario where the absent low-intensity peak is a structural element, rather than an artefact.

Fourier-domain Q-OCT is also very competitive with regards to the signal acquisition time. The summary of the total acquisition times for all objects imaged with the Fourier-domain Q-OCT is presented in Table~\ref{tab:time}, with the corresponding acquisition times for Time-domain Q-OCT and conventional OCT included for reference. The Fourier-domain approach outperformed the time-domain approach: the more extended the imaged object is, the bigger the difference in the acquisition time. This is due to the fact that bulkier objects require a longer axial scan of the reference arm stage, resulting in the need to acquire more data points. In the case of a Fourier-domain signal, only a single joint spectrum is required. Although the acquisition times of Fourier-domain Q-OCT are superior to the ones in Time-domain Q-OCT, they are nowhere near the times achieved in conventional OCT, even when conventional OCT is performed with non-OCT-specific equipment, such as Optical Spectrum Analyser as is the case here.

% imaging depth limits
As is generally the case in all Fourier-domain approaches, the imaging depth of Fourier-domain Q-OCT is restricted by the spectral resolution of the detection. The 5-kilometre-long fibre spools allow only 0.24~mm depth to be visualised, \figref{fig:fall-off}, resulting in barely visualising the back surface of the 260-$\mu$m-thick plastic (\figref{fig:plastic}) and failure to visualise the bottom-most surface in the glass stack (\figref{fig:2glass}). Longer fibre spools would increase the spectral resolution and extend the imaging depth, but at the same time, they would naturally attenuate the light more and consequently, require longer acquisition times.

There are also technical aspects to consider related to setting up and operating a Q-OCT system. Because the intensity of the quantum light is extremely low, one uses the classical laser light source to built and optimise the experimental system, ending up building a traditional OCT system either way. Similarly, due to long acquisition times, Q-OCT cannot be performed in real time. Because of that, the optimum position of the object in the object arm is found by switching to the classical light and maximising the intensity of light reflected from it. Consequently, conventional OCT remains to be an indispensible reference when performing the Q-OCT experiment.

One aspect of Fourier-domain Q-OCT which was shown here and which necessitates a separate broader study is the influence of the chromatic dispersion effects introduced by the dispersive elements in the detection and the spectrally broadband pump. A better understanding of the connection between these effects and the behaviour of both the structural and artefact elements in the 2D Fourier transform will result in better dispersion-removing algorithms for Fourier-domain Q-OCT. Also, with a proper model of the experimental system, more efficient algorithms for artefact removal could be devised, including - as mentioned before - ones based on Machine Learning.

Another aspect worth exploring is further speeding up the measurement. What data acquisition parallelisation schemes could be employed to reduce the current record of 4 minutes for non-mirror objects (reported in this article)? Is there a scheme that will allow the world's first fully non-invasive Q-OCT imaging of biological specimens? Or should we rather expect to achieve this when sufficient progress has been made in quantum light generation and detection?

\bibliography{main}

\begin{thebibliography}{23}%
\makeatletter
\providecommand \@ifxundefined [1]{%
 \@ifx{#1\undefined}
}%
\providecommand \@ifnum [1]{%
 \ifnum #1\expandafter \@firstoftwo
 \else \expandafter \@secondoftwo
 \fi
}%
\providecommand \@ifx [1]{%
 \ifx #1\expandafter \@firstoftwo
 \else \expandafter \@secondoftwo
 \fi
}%
\providecommand \natexlab [1]{#1}%
\providecommand \enquote  [1]{``#1''}%
\providecommand \bibnamefont  [1]{#1}%
\providecommand \bibfnamefont [1]{#1}%
\providecommand \citenamefont [1]{#1}%
\providecommand \href@noop [0]{\@secondoftwo}%
\providecommand \href [0]{\begingroup \@sanitize@url \@href}%
\providecommand \@href[1]{\@@startlink{#1}\@@href}%
\providecommand \@@href[1]{\endgroup#1\@@endlink}%
\providecommand \@sanitize@url [0]{\catcode `\\12\catcode `\$12\catcode `\&12\catcode `\#12\catcode `\^12\catcode `\_12\catcode `\%12\relax}%
\providecommand \@@startlink[1]{}%
\providecommand \@@endlink[0]{}%
\providecommand \url  [0]{\begingroup\@sanitize@url \@url }%
\providecommand \@url [1]{\endgroup\@href {#1}{\urlprefix }}%
\providecommand \urlprefix  [0]{URL }%
\providecommand \Eprint [0]{\href }%
\providecommand \doibase [0]{http://dx.doi.org/}%
\providecommand \selectlanguage [0]{\@gobble}%
\providecommand \bibinfo  [0]{\@secondoftwo}%
\providecommand \bibfield  [0]{\@secondoftwo}%
\providecommand \translation [1]{[#1]}%
\providecommand \BibitemOpen [0]{}%
\providecommand \bibitemStop [0]{}%
\providecommand \bibitemNoStop [0]{.\EOS\space}%
\providecommand \EOS [0]{\spacefactor3000\relax}%
\providecommand \BibitemShut  [1]{\csname bibitem#1\endcsname}%
\let\auto@bib@innerbib\@empty
%</preamble>
\bibitem [{\citenamefont {Abouraddy}\ \emph {et~al.}(2002)\citenamefont {Abouraddy}, \citenamefont {Nasr}, \citenamefont {Saleh}, \citenamefont {Sergienko},\ and\ \citenamefont {Teich}}]{abouraddy2002quantum}%
  \BibitemOpen
  \bibfield  {author} {\bibinfo {author} {\bibfnamefont {A.~F.}\ \bibnamefont {Abouraddy}}, \bibinfo {author} {\bibfnamefont {M.~B.}\ \bibnamefont {Nasr}}, \bibinfo {author} {\bibfnamefont {B.~E.}\ \bibnamefont {Saleh}}, \bibinfo {author} {\bibfnamefont {A.~V.}\ \bibnamefont {Sergienko}}, \ and\ \bibinfo {author} {\bibfnamefont {M.~C.}\ \bibnamefont {Teich}},\ }\href@noop {} {\bibfield  {journal} {\bibinfo  {journal} {Physical Review A}\ }\textbf {\bibinfo {volume} {65}},\ \bibinfo {pages} {053817} (\bibinfo {year} {2002})}\BibitemShut {NoStop}%
\bibitem [{\citenamefont {Nasr}\ \emph {et~al.}(2003)\citenamefont {Nasr}, \citenamefont {Saleh}, \citenamefont {Sergienko},\ and\ \citenamefont {Teich}}]{nasr2003demonstration}%
  \BibitemOpen
  \bibfield  {author} {\bibinfo {author} {\bibfnamefont {M.~B.}\ \bibnamefont {Nasr}}, \bibinfo {author} {\bibfnamefont {B.~E.}\ \bibnamefont {Saleh}}, \bibinfo {author} {\bibfnamefont {A.~V.}\ \bibnamefont {Sergienko}}, \ and\ \bibinfo {author} {\bibfnamefont {M.~C.}\ \bibnamefont {Teich}},\ }\href@noop {} {\bibfield  {journal} {\bibinfo  {journal} {Physical review letters}\ }\textbf {\bibinfo {volume} {91}},\ \bibinfo {pages} {083601} (\bibinfo {year} {2003})}\BibitemShut {NoStop}%
\bibitem [{\citenamefont {Kolenderska}\ \emph {et~al.}(2020{\natexlab{a}})\citenamefont {Kolenderska}, \citenamefont {Vanholsbeeck},\ and\ \citenamefont {Kolenderski}}]{kolenderska2020quantum}%
  \BibitemOpen
  \bibfield  {author} {\bibinfo {author} {\bibfnamefont {S.~M.}\ \bibnamefont {Kolenderska}}, \bibinfo {author} {\bibfnamefont {F.}~\bibnamefont {Vanholsbeeck}}, \ and\ \bibinfo {author} {\bibfnamefont {P.}~\bibnamefont {Kolenderski}},\ }\href@noop {} {\bibfield  {journal} {\bibinfo  {journal} {Optics Letters}\ }\textbf {\bibinfo {volume} {45}},\ \bibinfo {pages} {3443} (\bibinfo {year} {2020}{\natexlab{a}})}\BibitemShut {NoStop}%
\bibitem [{\citenamefont {Hong}\ \emph {et~al.}(1987)\citenamefont {Hong}, \citenamefont {Ou},\ and\ \citenamefont {Mandel}}]{hong1987measurement}%
  \BibitemOpen
  \bibfield  {author} {\bibinfo {author} {\bibfnamefont {C.-K.}\ \bibnamefont {Hong}}, \bibinfo {author} {\bibfnamefont {Z.-Y.}\ \bibnamefont {Ou}}, \ and\ \bibinfo {author} {\bibfnamefont {L.}~\bibnamefont {Mandel}},\ }\href@noop {} {\bibfield  {journal} {\bibinfo  {journal} {Physical review letters}\ }\textbf {\bibinfo {volume} {59}},\ \bibinfo {pages} {2044} (\bibinfo {year} {1987})}\BibitemShut {NoStop}%
\bibitem [{\citenamefont {Hayama}\ \emph {et~al.}(2022)\citenamefont {Hayama}, \citenamefont {Cao}, \citenamefont {Okamoto}, \citenamefont {Suezawa}, \citenamefont {Okano},\ and\ \citenamefont {Takeuchi}}]{hayama2022high}%
  \BibitemOpen
  \bibfield  {author} {\bibinfo {author} {\bibfnamefont {K.}~\bibnamefont {Hayama}}, \bibinfo {author} {\bibfnamefont {B.}~\bibnamefont {Cao}}, \bibinfo {author} {\bibfnamefont {R.}~\bibnamefont {Okamoto}}, \bibinfo {author} {\bibfnamefont {S.}~\bibnamefont {Suezawa}}, \bibinfo {author} {\bibfnamefont {M.}~\bibnamefont {Okano}}, \ and\ \bibinfo {author} {\bibfnamefont {S.}~\bibnamefont {Takeuchi}},\ }\href@noop {} {\bibfield  {journal} {\bibinfo  {journal} {Optics Letters}\ }\textbf {\bibinfo {volume} {47}},\ \bibinfo {pages} {4949} (\bibinfo {year} {2022})}\BibitemShut {NoStop}%
\bibitem [{\citenamefont {Yepiz-Graciano}\ \emph {et~al.}(2022)\citenamefont {Yepiz-Graciano}, \citenamefont {Ibarra-Borja}, \citenamefont {Ram{\'\i}rez~Alarc{\'o}n}, \citenamefont {Guti{\'e}rrez-Torres}, \citenamefont {Cruz-Ram{\'\i}rez}, \citenamefont {Lopez-Mago},\ and\ \citenamefont {U’Ren}}]{yepiz2022quantum}%
  \BibitemOpen
  \bibfield  {author} {\bibinfo {author} {\bibfnamefont {P.}~\bibnamefont {Yepiz-Graciano}}, \bibinfo {author} {\bibfnamefont {Z.}~\bibnamefont {Ibarra-Borja}}, \bibinfo {author} {\bibfnamefont {R.}~\bibnamefont {Ram{\'\i}rez~Alarc{\'o}n}}, \bibinfo {author} {\bibfnamefont {G.}~\bibnamefont {Guti{\'e}rrez-Torres}}, \bibinfo {author} {\bibfnamefont {H.}~\bibnamefont {Cruz-Ram{\'\i}rez}}, \bibinfo {author} {\bibfnamefont {D.}~\bibnamefont {Lopez-Mago}}, \ and\ \bibinfo {author} {\bibfnamefont {A.~B.}\ \bibnamefont {U’Ren}},\ }\href@noop {} {\bibfield  {journal} {\bibinfo  {journal} {Physical Review Applied}\ }\textbf {\bibinfo {volume} {18}},\ \bibinfo {pages} {034060} (\bibinfo {year} {2022})}\BibitemShut {NoStop}%
\bibitem [{\citenamefont {Sukharenko}\ \emph {et~al.}(2021)\citenamefont {Sukharenko}, \citenamefont {Bikorimana},\ and\ \citenamefont {Dorsinville}}]{sukharenko2021birefringence}%
  \BibitemOpen
  \bibfield  {author} {\bibinfo {author} {\bibfnamefont {V.}~\bibnamefont {Sukharenko}}, \bibinfo {author} {\bibfnamefont {S.}~\bibnamefont {Bikorimana}}, \ and\ \bibinfo {author} {\bibfnamefont {R.}~\bibnamefont {Dorsinville}},\ }\href@noop {} {\bibfield  {journal} {\bibinfo  {journal} {Optics Letters}\ }\textbf {\bibinfo {volume} {46}},\ \bibinfo {pages} {2799} (\bibinfo {year} {2021})}\BibitemShut {NoStop}%
\bibitem [{\citenamefont {Ibarra-Borja}\ \emph {et~al.}(2019)\citenamefont {Ibarra-Borja}, \citenamefont {Sevilla-Guti{\'e}rrez}, \citenamefont {Ram{\'\i}rez-Alarc{\'o}n}, \citenamefont {Cruz-Ram{\'\i}rez},\ and\ \citenamefont {U’Ren}}]{ibarra2019experimental}%
  \BibitemOpen
  \bibfield  {author} {\bibinfo {author} {\bibfnamefont {Z.}~\bibnamefont {Ibarra-Borja}}, \bibinfo {author} {\bibfnamefont {C.}~\bibnamefont {Sevilla-Guti{\'e}rrez}}, \bibinfo {author} {\bibfnamefont {R.}~\bibnamefont {Ram{\'\i}rez-Alarc{\'o}n}}, \bibinfo {author} {\bibfnamefont {H.}~\bibnamefont {Cruz-Ram{\'\i}rez}}, \ and\ \bibinfo {author} {\bibfnamefont {A.~B.}\ \bibnamefont {U’Ren}},\ }\href@noop {} {\bibfield  {journal} {\bibinfo  {journal} {Photonics Research}\ }\textbf {\bibinfo {volume} {8}},\ \bibinfo {pages} {51} (\bibinfo {year} {2019})}\BibitemShut {NoStop}%
\bibitem [{\citenamefont {Okano}\ \emph {et~al.}(2015)\citenamefont {Okano}, \citenamefont {Lim}, \citenamefont {Okamoto}, \citenamefont {Nishizawa}, \citenamefont {Kurimura},\ and\ \citenamefont {Takeuchi}}]{okano20150}%
  \BibitemOpen
  \bibfield  {author} {\bibinfo {author} {\bibfnamefont {M.}~\bibnamefont {Okano}}, \bibinfo {author} {\bibfnamefont {H.~H.}\ \bibnamefont {Lim}}, \bibinfo {author} {\bibfnamefont {R.}~\bibnamefont {Okamoto}}, \bibinfo {author} {\bibfnamefont {N.}~\bibnamefont {Nishizawa}}, \bibinfo {author} {\bibfnamefont {S.}~\bibnamefont {Kurimura}}, \ and\ \bibinfo {author} {\bibfnamefont {S.}~\bibnamefont {Takeuchi}},\ }\href@noop {} {\bibfield  {journal} {\bibinfo  {journal} {Scientific reports}\ }\textbf {\bibinfo {volume} {5}},\ \bibinfo {pages} {18042} (\bibinfo {year} {2015})}\BibitemShut {NoStop}%
\bibitem [{\citenamefont {Lopez-Mago}\ and\ \citenamefont {Novotny}(2012)}]{lopez2012quantum}%
  \BibitemOpen
  \bibfield  {author} {\bibinfo {author} {\bibfnamefont {D.}~\bibnamefont {Lopez-Mago}}\ and\ \bibinfo {author} {\bibfnamefont {L.}~\bibnamefont {Novotny}},\ }in\ \href@noop {} {\emph {\bibinfo {booktitle} {Frontiers in Optics}}}\ (\bibinfo {organization} {Optica Publishing Group},\ \bibinfo {year} {2012})\ pp.\ \bibinfo {pages} {FTh4E--3}\BibitemShut {NoStop}%
\bibitem [{\citenamefont {Nasr}\ \emph {et~al.}(2009)\citenamefont {Nasr}, \citenamefont {Goode}, \citenamefont {Nguyen}, \citenamefont {Rong}, \citenamefont {Yang}, \citenamefont {Reinhard}, \citenamefont {Saleh},\ and\ \citenamefont {Teich}}]{nasr2009quantum}%
  \BibitemOpen
  \bibfield  {author} {\bibinfo {author} {\bibfnamefont {M.~B.}\ \bibnamefont {Nasr}}, \bibinfo {author} {\bibfnamefont {D.~P.}\ \bibnamefont {Goode}}, \bibinfo {author} {\bibfnamefont {N.}~\bibnamefont {Nguyen}}, \bibinfo {author} {\bibfnamefont {G.}~\bibnamefont {Rong}}, \bibinfo {author} {\bibfnamefont {L.}~\bibnamefont {Yang}}, \bibinfo {author} {\bibfnamefont {B.~M.}\ \bibnamefont {Reinhard}}, \bibinfo {author} {\bibfnamefont {B.~E.}\ \bibnamefont {Saleh}}, \ and\ \bibinfo {author} {\bibfnamefont {M.~C.}\ \bibnamefont {Teich}},\ }\href@noop {} {\bibfield  {journal} {\bibinfo  {journal} {Optics Communications}\ }\textbf {\bibinfo {volume} {282}},\ \bibinfo {pages} {1154} (\bibinfo {year} {2009})}\BibitemShut {NoStop}%
\bibitem [{\citenamefont {Nasr}\ \emph {et~al.}(2004)\citenamefont {Nasr}, \citenamefont {Saleh}, \citenamefont {Sergienko},\ and\ \citenamefont {Teich}}]{nasr2004dispersion}%
  \BibitemOpen
  \bibfield  {author} {\bibinfo {author} {\bibfnamefont {M.~B.}\ \bibnamefont {Nasr}}, \bibinfo {author} {\bibfnamefont {B.~E.}\ \bibnamefont {Saleh}}, \bibinfo {author} {\bibfnamefont {A.~V.}\ \bibnamefont {Sergienko}}, \ and\ \bibinfo {author} {\bibfnamefont {M.~C.}\ \bibnamefont {Teich}},\ }\href@noop {} {\bibfield  {journal} {\bibinfo  {journal} {Optics express}\ }\textbf {\bibinfo {volume} {12}},\ \bibinfo {pages} {1353} (\bibinfo {year} {2004})}\BibitemShut {NoStop}%
\bibitem [{\citenamefont {Yepiz-Graciano}\ \emph {et~al.}(2020)\citenamefont {Yepiz-Graciano}, \citenamefont {Mart{\'\i}nez}, \citenamefont {Lopez-Mago}, \citenamefont {Cruz-Ramirez},\ and\ \citenamefont {U’Ren}}]{yepiz2020spectrally}%
  \BibitemOpen
  \bibfield  {author} {\bibinfo {author} {\bibfnamefont {P.}~\bibnamefont {Yepiz-Graciano}}, \bibinfo {author} {\bibfnamefont {A.~M.~A.}\ \bibnamefont {Mart{\'\i}nez}}, \bibinfo {author} {\bibfnamefont {D.}~\bibnamefont {Lopez-Mago}}, \bibinfo {author} {\bibfnamefont {H.}~\bibnamefont {Cruz-Ramirez}}, \ and\ \bibinfo {author} {\bibfnamefont {A.~B.}\ \bibnamefont {U’Ren}},\ }\href@noop {} {\bibfield  {journal} {\bibinfo  {journal} {Photonics Research}\ }\textbf {\bibinfo {volume} {8}},\ \bibinfo {pages} {1023} (\bibinfo {year} {2020})}\BibitemShut {NoStop}%
\bibitem [{\citenamefont {Graciano}\ \emph {et~al.}(2019)\citenamefont {Graciano}, \citenamefont {Mart{\'\i}nez}, \citenamefont {Lopez-Mago}, \citenamefont {Castro-Olvera}, \citenamefont {Rosete-Aguilar}, \citenamefont {Gardu{\~n}o-Mej{\'\i}a}, \citenamefont {Alarc{\'o}n}, \citenamefont {Ram{\'\i}rez},\ and\ \citenamefont {U’Ren}}]{graciano2019interference}%
  \BibitemOpen
  \bibfield  {author} {\bibinfo {author} {\bibfnamefont {P.~Y.}\ \bibnamefont {Graciano}}, \bibinfo {author} {\bibfnamefont {A.~M.~A.}\ \bibnamefont {Mart{\'\i}nez}}, \bibinfo {author} {\bibfnamefont {D.}~\bibnamefont {Lopez-Mago}}, \bibinfo {author} {\bibfnamefont {G.}~\bibnamefont {Castro-Olvera}}, \bibinfo {author} {\bibfnamefont {M.}~\bibnamefont {Rosete-Aguilar}}, \bibinfo {author} {\bibfnamefont {J.}~\bibnamefont {Gardu{\~n}o-Mej{\'\i}a}}, \bibinfo {author} {\bibfnamefont {R.~R.}\ \bibnamefont {Alarc{\'o}n}}, \bibinfo {author} {\bibfnamefont {H.~C.}\ \bibnamefont {Ram{\'\i}rez}}, \ and\ \bibinfo {author} {\bibfnamefont {A.~B.}\ \bibnamefont {U’Ren}},\ }\href@noop {} {\bibfield  {journal} {\bibinfo  {journal} {Scientific Reports}\ }\textbf {\bibinfo {volume} {9}},\ \bibinfo {pages} {1} (\bibinfo {year} {2019})}\BibitemShut {NoStop}%
\bibitem [{\citenamefont {Teich}\ \emph {et~al.}(2012)\citenamefont {Teich}, \citenamefont {Saleh}, \citenamefont {Wong},\ and\ \citenamefont {Shapiro}}]{teich2012variations}%
  \BibitemOpen
  \bibfield  {author} {\bibinfo {author} {\bibfnamefont {M.~C.}\ \bibnamefont {Teich}}, \bibinfo {author} {\bibfnamefont {B.~E.}\ \bibnamefont {Saleh}}, \bibinfo {author} {\bibfnamefont {F.~N.}\ \bibnamefont {Wong}}, \ and\ \bibinfo {author} {\bibfnamefont {J.~H.}\ \bibnamefont {Shapiro}},\ }\href@noop {} {\bibfield  {journal} {\bibinfo  {journal} {Quantum Information Processing}\ }\textbf {\bibinfo {volume} {11}},\ \bibinfo {pages} {903} (\bibinfo {year} {2012})}\BibitemShut {NoStop}%
\bibitem [{\citenamefont {Kolenderska}\ \emph {et~al.}(2019)\citenamefont {Kolenderska}, \citenamefont {Vanholsbeeck},\ and\ \citenamefont {Kolenderski}}]{kolenderska2019}%
  \BibitemOpen
  \bibfield  {author} {\bibinfo {author} {\bibfnamefont {S.~M.}\ \bibnamefont {Kolenderska}}, \bibinfo {author} {\bibfnamefont {F.}~\bibnamefont {Vanholsbeeck}}, \ and\ \bibinfo {author} {\bibfnamefont {P.}~\bibnamefont {Kolenderski}},\ }in\ \href@noop {} {\emph {\bibinfo {booktitle} {European Conference on Biomedical Optics}}}\ (\bibinfo {organization} {Optica Publishing Group},\ \bibinfo {year} {2019})\ pp.\ \bibinfo {pages} {11078--30}\BibitemShut {NoStop}%
\bibitem [{\citenamefont {Kolenderska}\ \emph {et~al.}(2020{\natexlab{b}})\citenamefont {Kolenderska}, \citenamefont {Vanholsbeeck},\ and\ \citenamefont {Kolenderski}}]{kolenderska2020fourier}%
  \BibitemOpen
  \bibfield  {author} {\bibinfo {author} {\bibfnamefont {S.~M.}\ \bibnamefont {Kolenderska}}, \bibinfo {author} {\bibfnamefont {F.}~\bibnamefont {Vanholsbeeck}}, \ and\ \bibinfo {author} {\bibfnamefont {P.}~\bibnamefont {Kolenderski}},\ }\href@noop {} {\bibfield  {journal} {\bibinfo  {journal} {Optics Express}\ }\textbf {\bibinfo {volume} {28}},\ \bibinfo {pages} {29576} (\bibinfo {year} {2020}{\natexlab{b}})}\BibitemShut {NoStop}%
\bibitem [{\citenamefont {Kolenderska}\ and\ \citenamefont {Szkulmowski}(2021)}]{kolenderska2021artefact}%
  \BibitemOpen
  \bibfield  {author} {\bibinfo {author} {\bibfnamefont {S.~M.}\ \bibnamefont {Kolenderska}}\ and\ \bibinfo {author} {\bibfnamefont {M.}~\bibnamefont {Szkulmowski}},\ }\href@noop {} {\bibfield  {journal} {\bibinfo  {journal} {Scientific reports}\ }\textbf {\bibinfo {volume} {11}},\ \bibinfo {pages} {18585} (\bibinfo {year} {2021})}\BibitemShut {NoStop}%
\bibitem [{\citenamefont {Okano}\ \emph {et~al.}(2013)\citenamefont {Okano}, \citenamefont {Okamoto}, \citenamefont {Tanaka}, \citenamefont {Ishida}, \citenamefont {Nishizawa},\ and\ \citenamefont {Takeuchi}}]{okano2013dispersion}%
  \BibitemOpen
  \bibfield  {author} {\bibinfo {author} {\bibfnamefont {M.}~\bibnamefont {Okano}}, \bibinfo {author} {\bibfnamefont {R.}~\bibnamefont {Okamoto}}, \bibinfo {author} {\bibfnamefont {A.}~\bibnamefont {Tanaka}}, \bibinfo {author} {\bibfnamefont {S.}~\bibnamefont {Ishida}}, \bibinfo {author} {\bibfnamefont {N.}~\bibnamefont {Nishizawa}}, \ and\ \bibinfo {author} {\bibfnamefont {S.}~\bibnamefont {Takeuchi}},\ }\href@noop {} {\bibfield  {journal} {\bibinfo  {journal} {Physical Review A—Atomic, Molecular, and Optical Physics}\ }\textbf {\bibinfo {volume} {88}},\ \bibinfo {pages} {043845} (\bibinfo {year} {2013})}\BibitemShut {NoStop}%
\bibitem [{\citenamefont {Hu}\ \emph {et~al.}(2007)\citenamefont {Hu}, \citenamefont {Pan},\ and\ \citenamefont {Rollins}}]{hu2007analytical}%
  \BibitemOpen
  \bibfield  {author} {\bibinfo {author} {\bibfnamefont {Z.}~\bibnamefont {Hu}}, \bibinfo {author} {\bibfnamefont {Y.}~\bibnamefont {Pan}}, \ and\ \bibinfo {author} {\bibfnamefont {A.~M.}\ \bibnamefont {Rollins}},\ }\href@noop {} {\bibfield  {journal} {\bibinfo  {journal} {Applied optics}\ }\textbf {\bibinfo {volume} {46}},\ \bibinfo {pages} {8499} (\bibinfo {year} {2007})}\BibitemShut {NoStop}%
\bibitem [{\citenamefont {Wang}\ and\ \citenamefont {Ding}(2008)}]{wang2008spectral}%
  \BibitemOpen
  \bibfield  {author} {\bibinfo {author} {\bibfnamefont {K.}~\bibnamefont {Wang}}\ and\ \bibinfo {author} {\bibfnamefont {Z.}~\bibnamefont {Ding}},\ }\href@noop {} {\bibfield  {journal} {\bibinfo  {journal} {Chinese Optics Letters}\ }\textbf {\bibinfo {volume} {6}},\ \bibinfo {pages} {902} (\bibinfo {year} {2008})}\BibitemShut {NoStop}%
\bibitem [{\citenamefont {Maliszewski}\ and\ \citenamefont {Kolenderska}(2021)}]{maliszewski2021artefact}%
  \BibitemOpen
  \bibfield  {author} {\bibinfo {author} {\bibfnamefont {K.~A.}\ \bibnamefont {Maliszewski}}\ and\ \bibinfo {author} {\bibfnamefont {S.~M.}\ \bibnamefont {Kolenderska}},\ }in\ \href@noop {} {\emph {\bibinfo {booktitle} {Optical Coherence Tomography and Coherence Domain Optical Methods in Biomedicine XXV}}},\ Vol.\ \bibinfo {volume} {11630}\ (\bibinfo {organization} {SPIE},\ \bibinfo {year} {2021})\ pp.\ \bibinfo {pages} {19--27}\BibitemShut {NoStop}%
\bibitem [{\citenamefont {Maliszewski}\ \emph {et~al.}(2022)\citenamefont {Maliszewski}, \citenamefont {Kolenderska},\ and\ \citenamefont {Vetrova}}]{maliszewski2022removing}%
  \BibitemOpen
  \bibfield  {author} {\bibinfo {author} {\bibfnamefont {K.~A.}\ \bibnamefont {Maliszewski}}, \bibinfo {author} {\bibfnamefont {S.~M.}\ \bibnamefont {Kolenderska}}, \ and\ \bibinfo {author} {\bibfnamefont {V.}~\bibnamefont {Vetrova}},\ }in\ \href@noop {} {\emph {\bibinfo {booktitle} {ICML 2022 2nd AI for Science Workshop}}}\ (\bibinfo {year} {2022})\BibitemShut {NoStop}%
\end{thebibliography}%

\section*{Acknowledgements}

SMK acknowledges New Zealand Ministry of Business, Innovation and Employment (MBIE) Smart Ideas funding (E7943). The authors acknowledge the financial support from Horizon Europe, the European Union's Framework Programme for Research and Innovation, SEQUOIA project, under Grant Agreement No. 101070062. 

\section*{Contributions}

SMK and PK conceived the idea, SK and FCVdB carried out the measurements. All authors reviewed the manuscript. 

\newpage

\begin{figure*}[htb]
    %\centering
    \includegraphics[clip=true, trim = 2cm 2cm 2cm 2cm, width=0.95\textwidth]{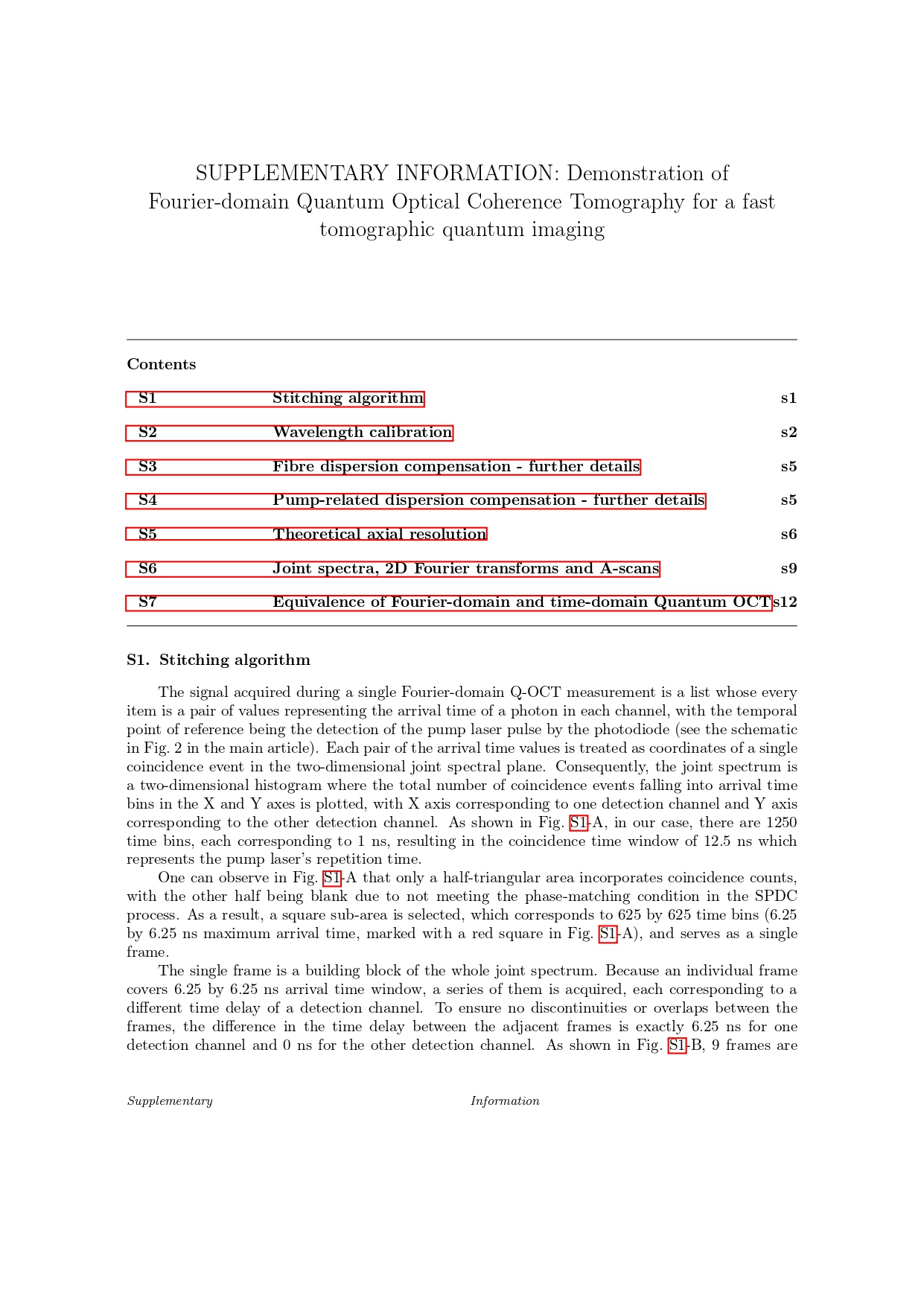}
\end{figure*}

\begin{figure*}[htb]
    %\centering
    \includegraphics[clip=true, trim = 2cm 2cm 2cm 2cm, width=0.95\textwidth]{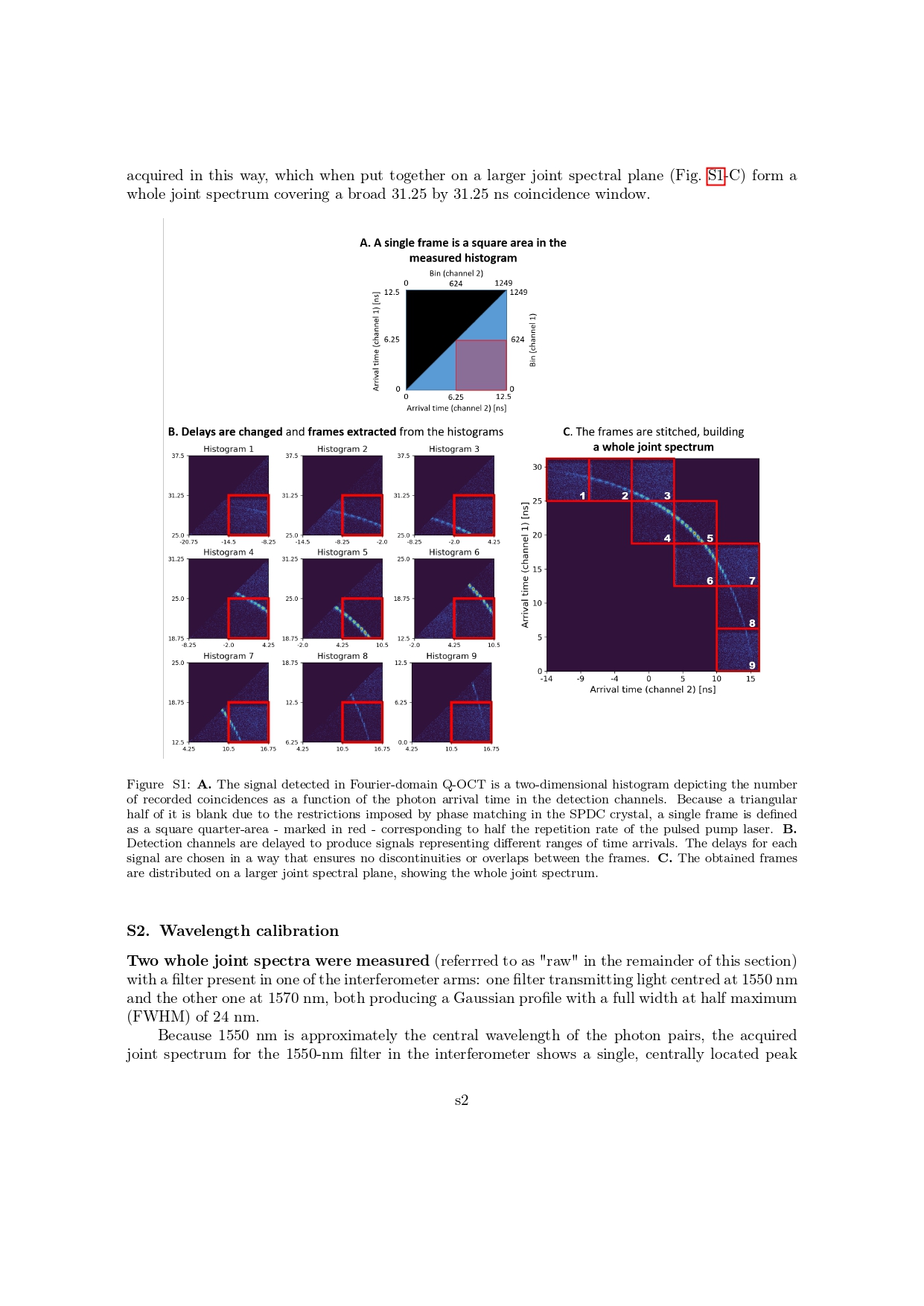}
\end{figure*}

\begin{figure*}[htb]
    %\centering
    \includegraphics[clip=true, trim = 2cm 2cm 2cm 2cm, width=0.95\textwidth]{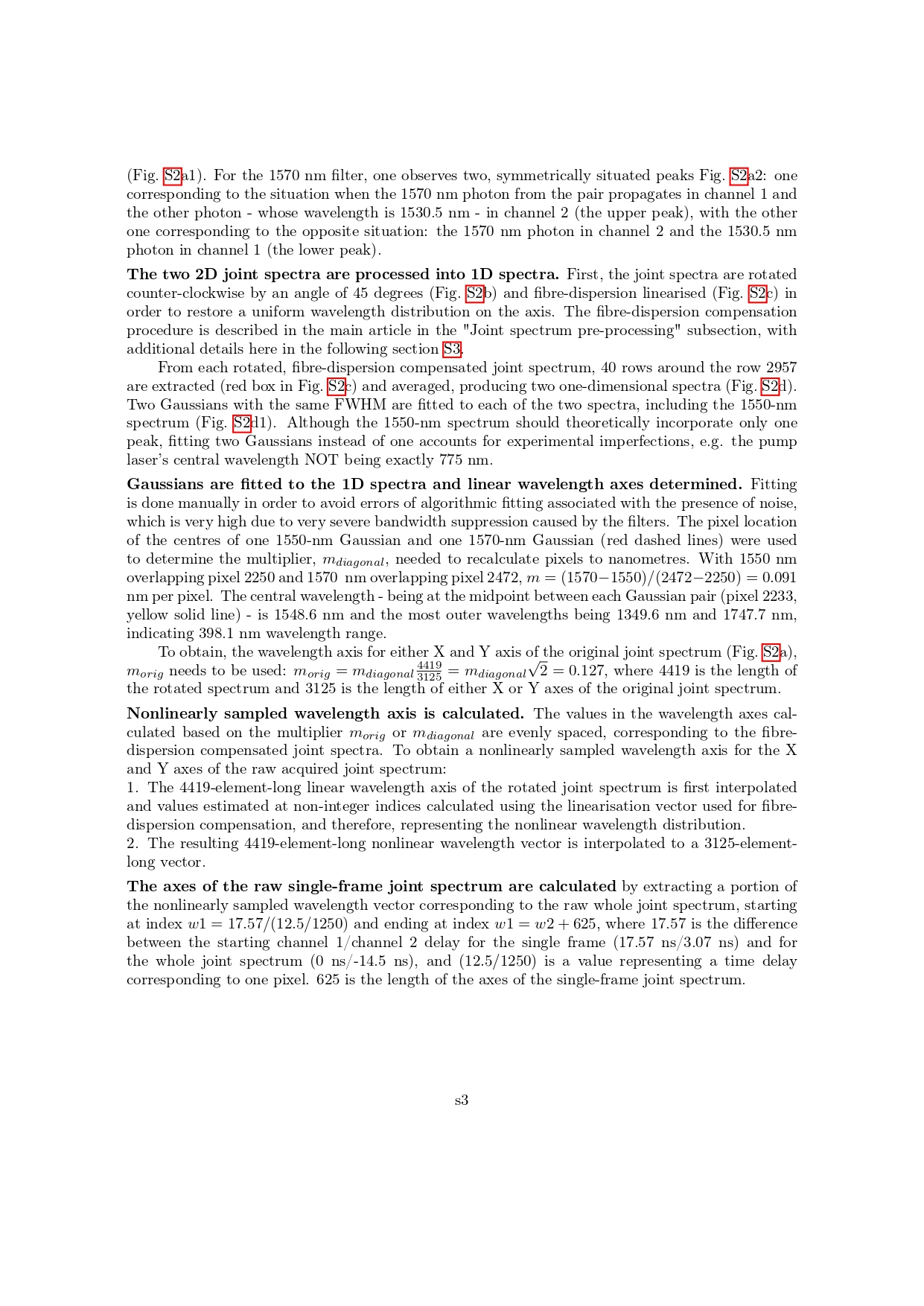}
\end{figure*}

\begin{figure*}[htb]
    %\centering
    \includegraphics[clip=true, trim = 2cm 2cm 2cm 2cm, width=0.95\textwidth]{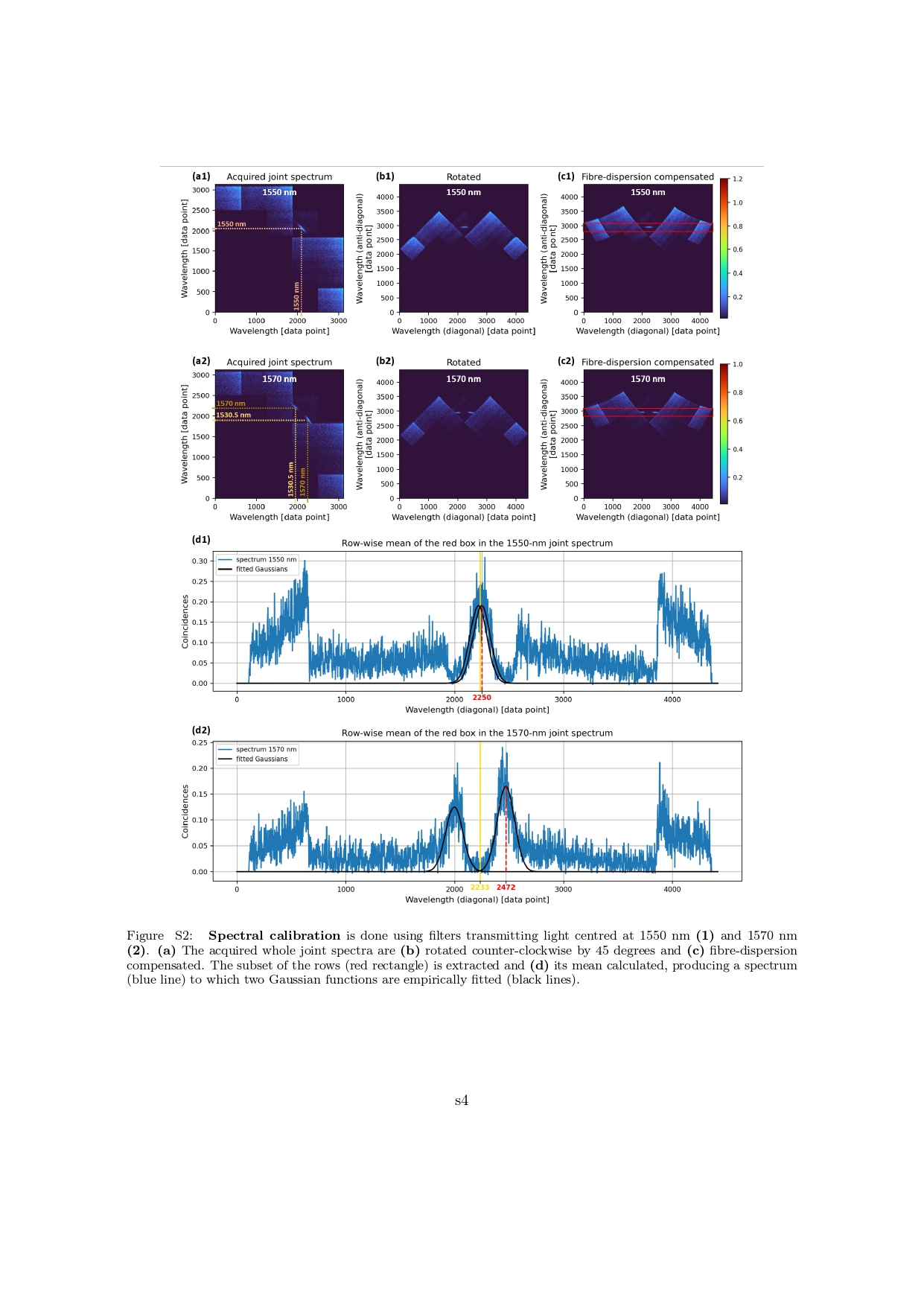}
\end{figure*}

\begin{figure*}[htb]
    %\centering
    \includegraphics[clip=true, trim = 2cm 2cm 2cm 2cm, width=0.95\textwidth]{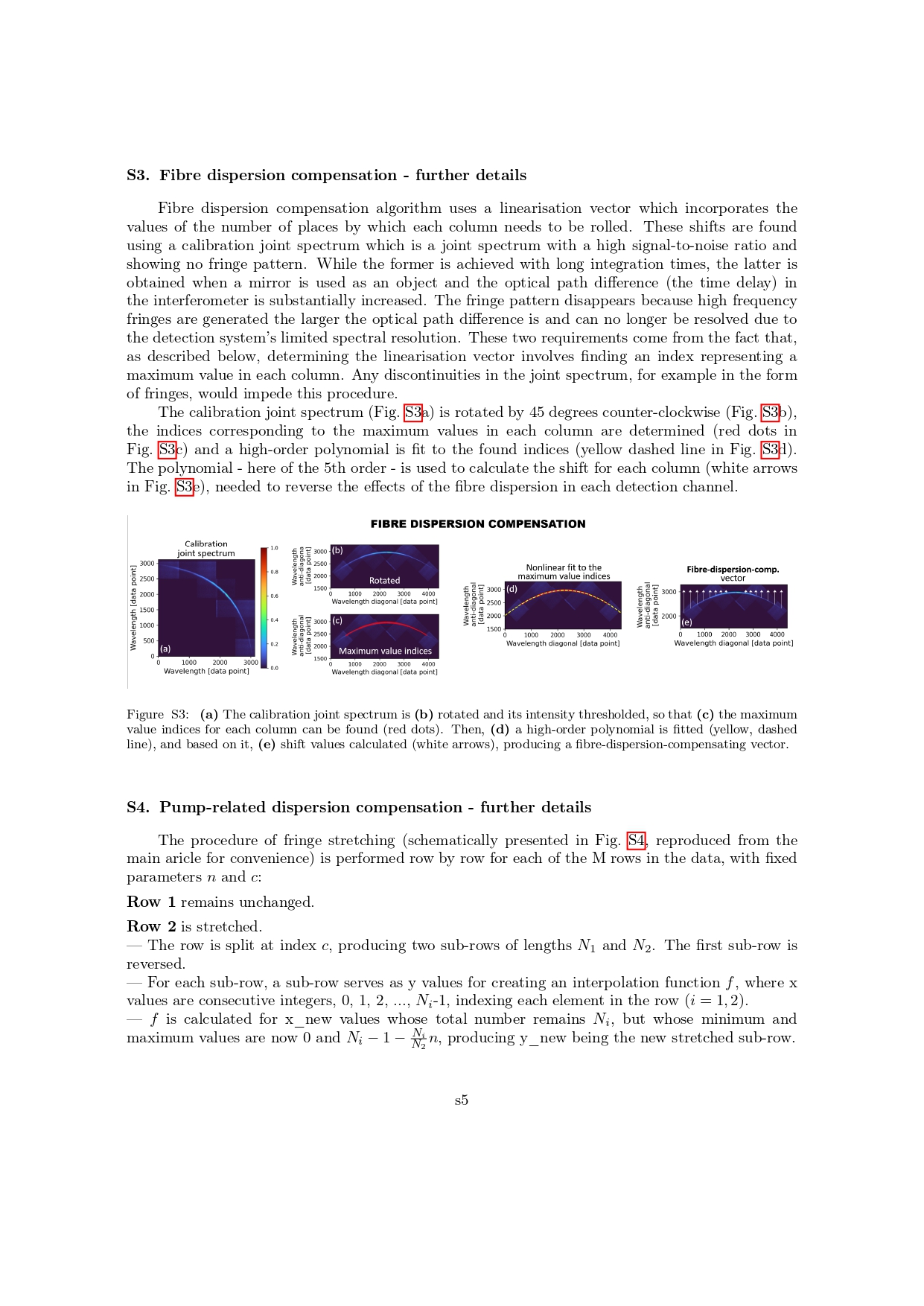}
\end{figure*}

\begin{figure*}[htb]
    %\centering
    \includegraphics[clip=true, trim = 2cm 2cm 2cm 2cm, width=0.95\textwidth]{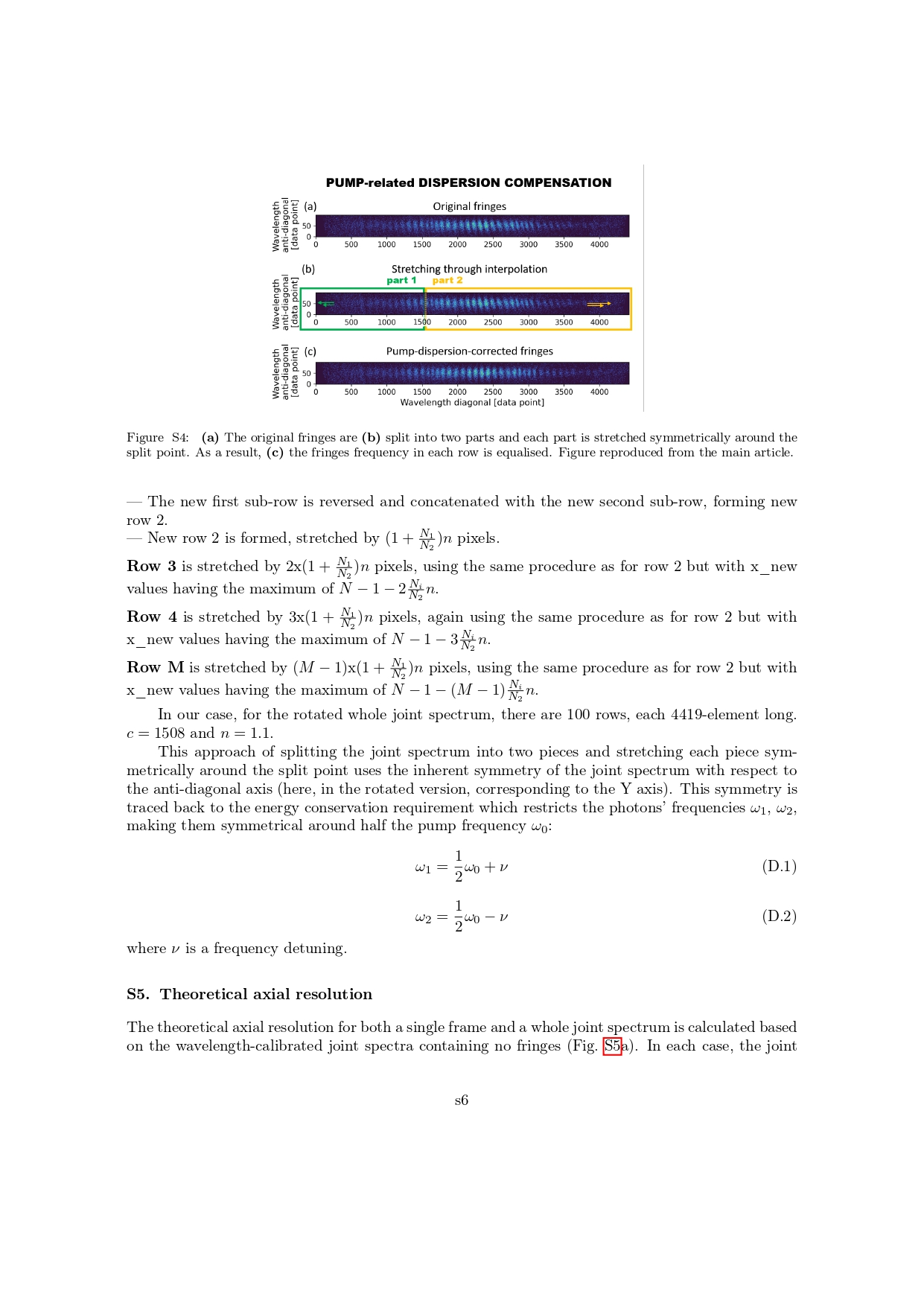}
\end{figure*}

\begin{figure*}[htb]
    %\centering
    \includegraphics[clip=true, trim = 2cm 2cm 2cm 2cm, width=0.95\textwidth]{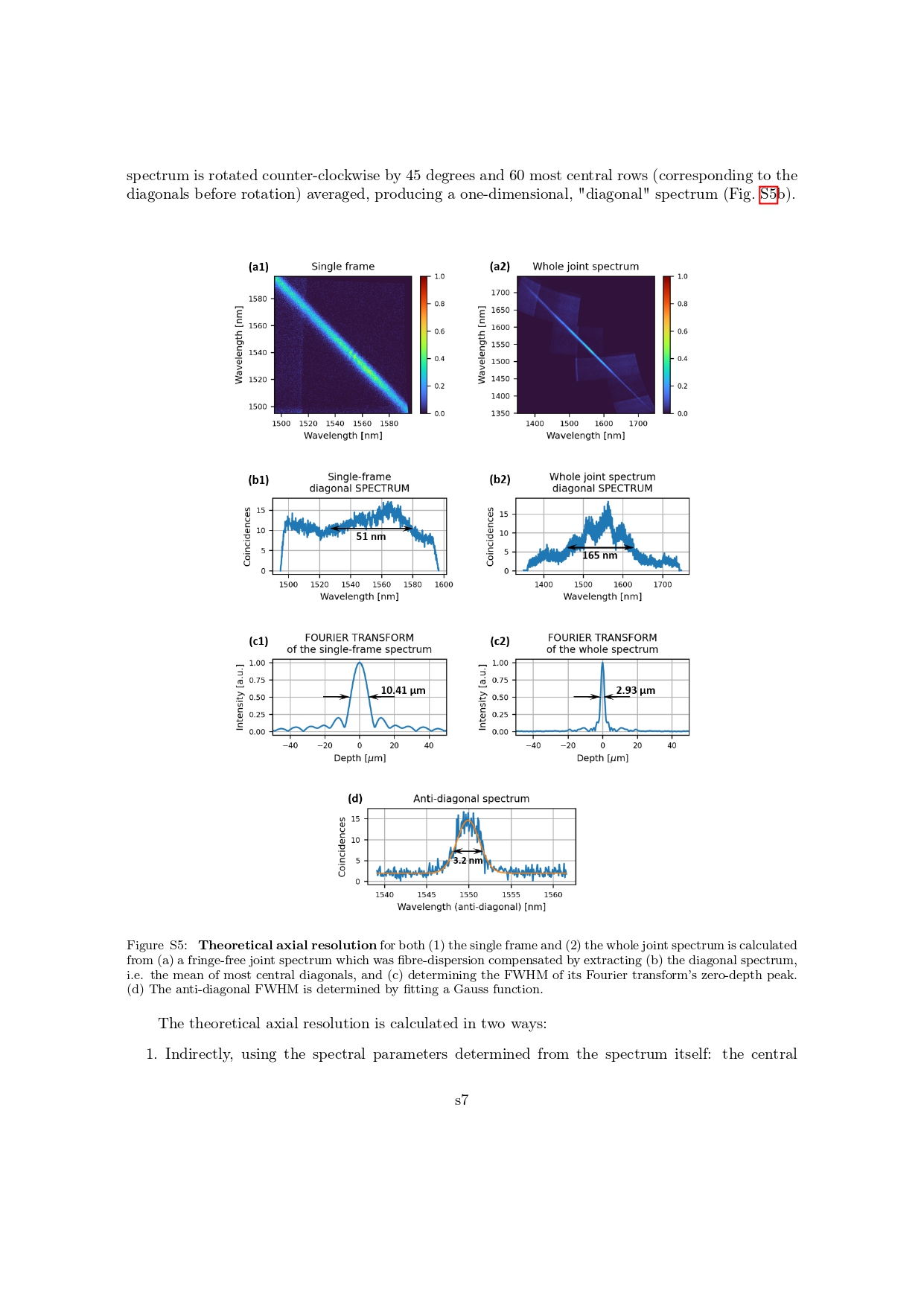}
\end{figure*}

\begin{figure*}[htb]
    %\centering
    \includegraphics[clip=true, trim = 2cm 2cm 2cm 2cm, width=0.95\textwidth]{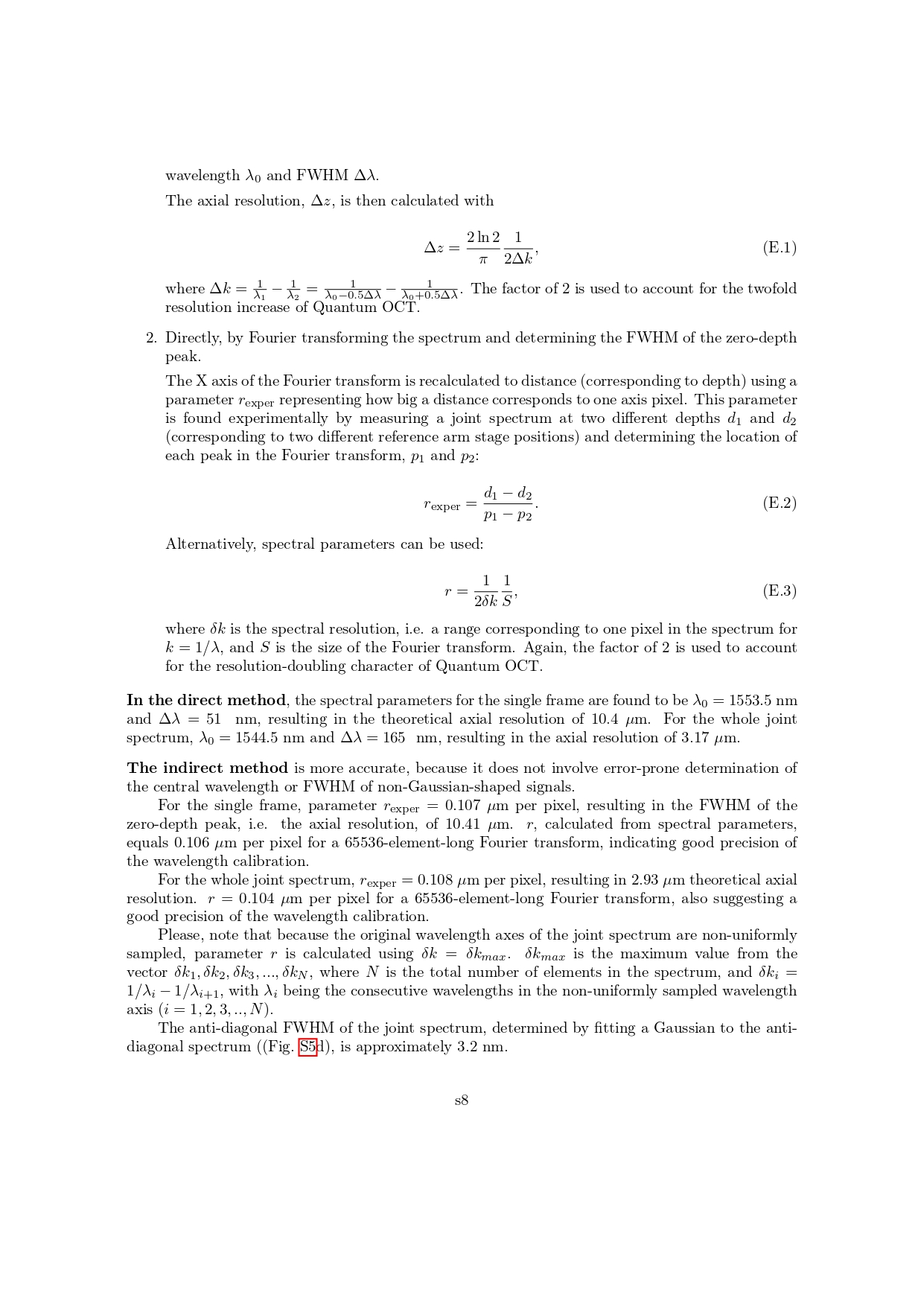}
\end{figure*}

\begin{figure*}[htb]
    %\centering
    \includegraphics[clip=true, trim = 2cm 2cm 2cm 2cm, width=0.95\textwidth]{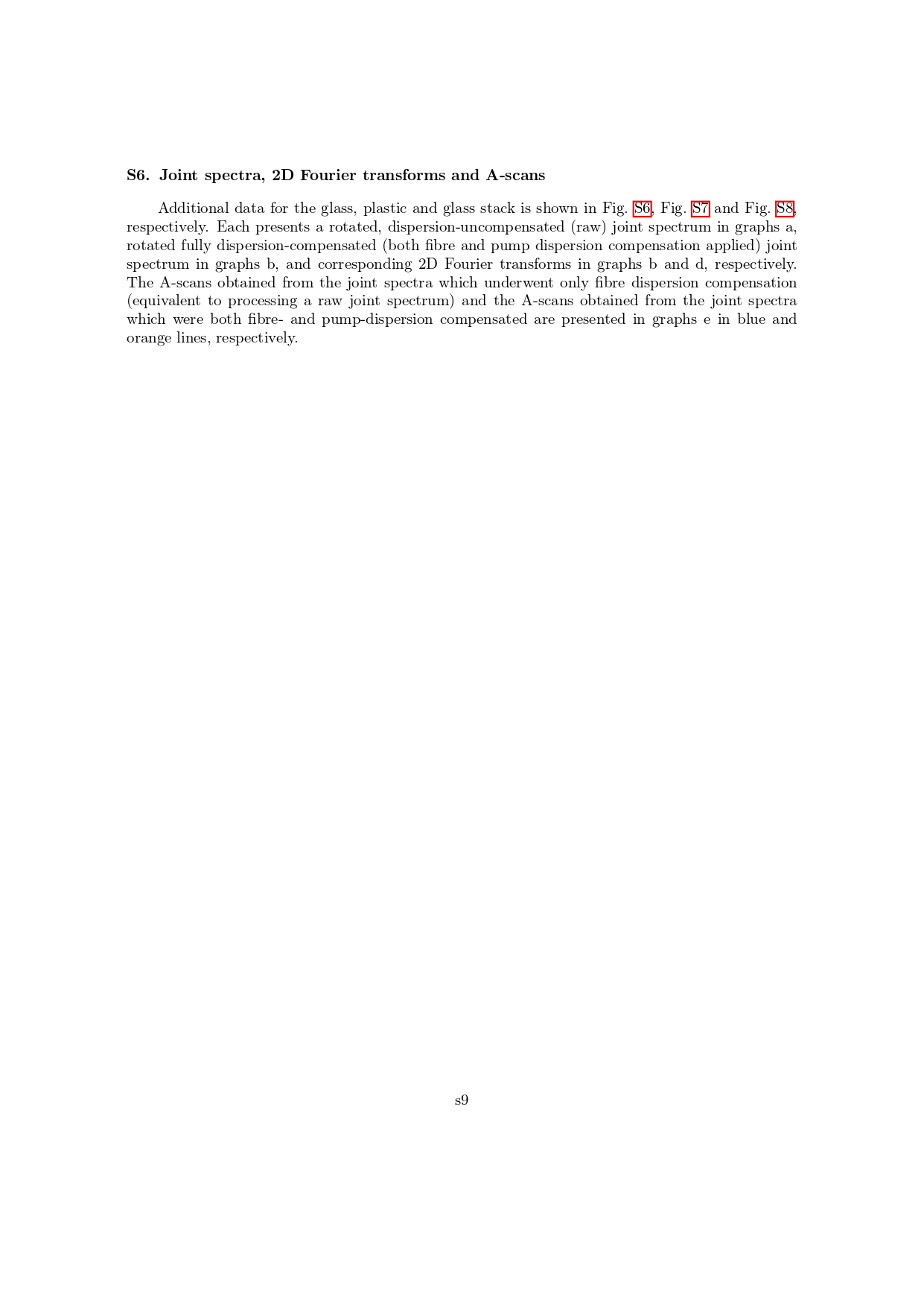}
\end{figure*}

\begin{figure*}[htb]
    %\centering
    \includegraphics[clip=true, trim = 2cm 2cm 2cm 2cm, width=0.95\textwidth]{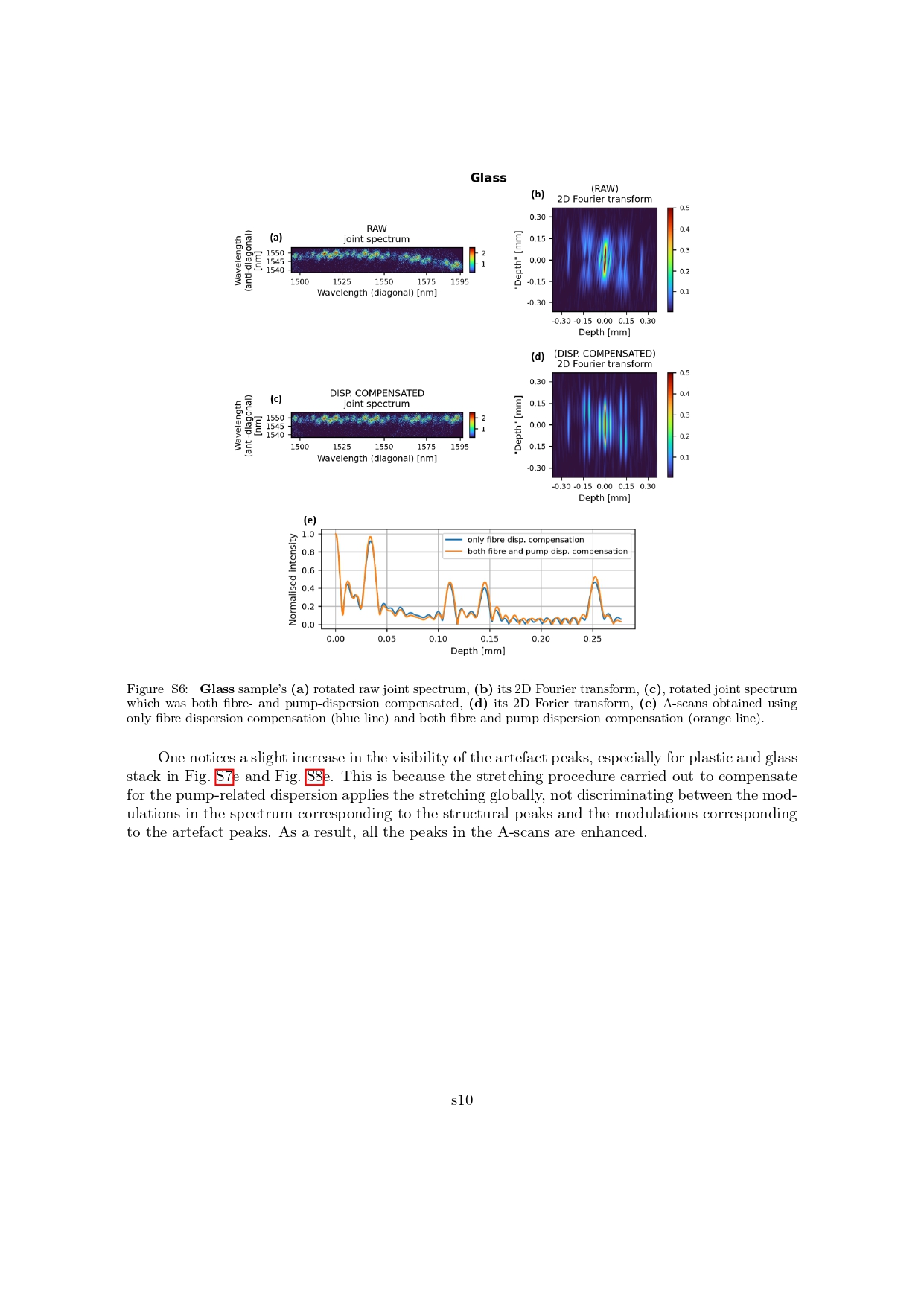}
\end{figure*}

\begin{figure*}[htb]
    %\centering
    \includegraphics[clip=true, trim = 2cm 2cm 2cm 2cm, width=0.95\textwidth]{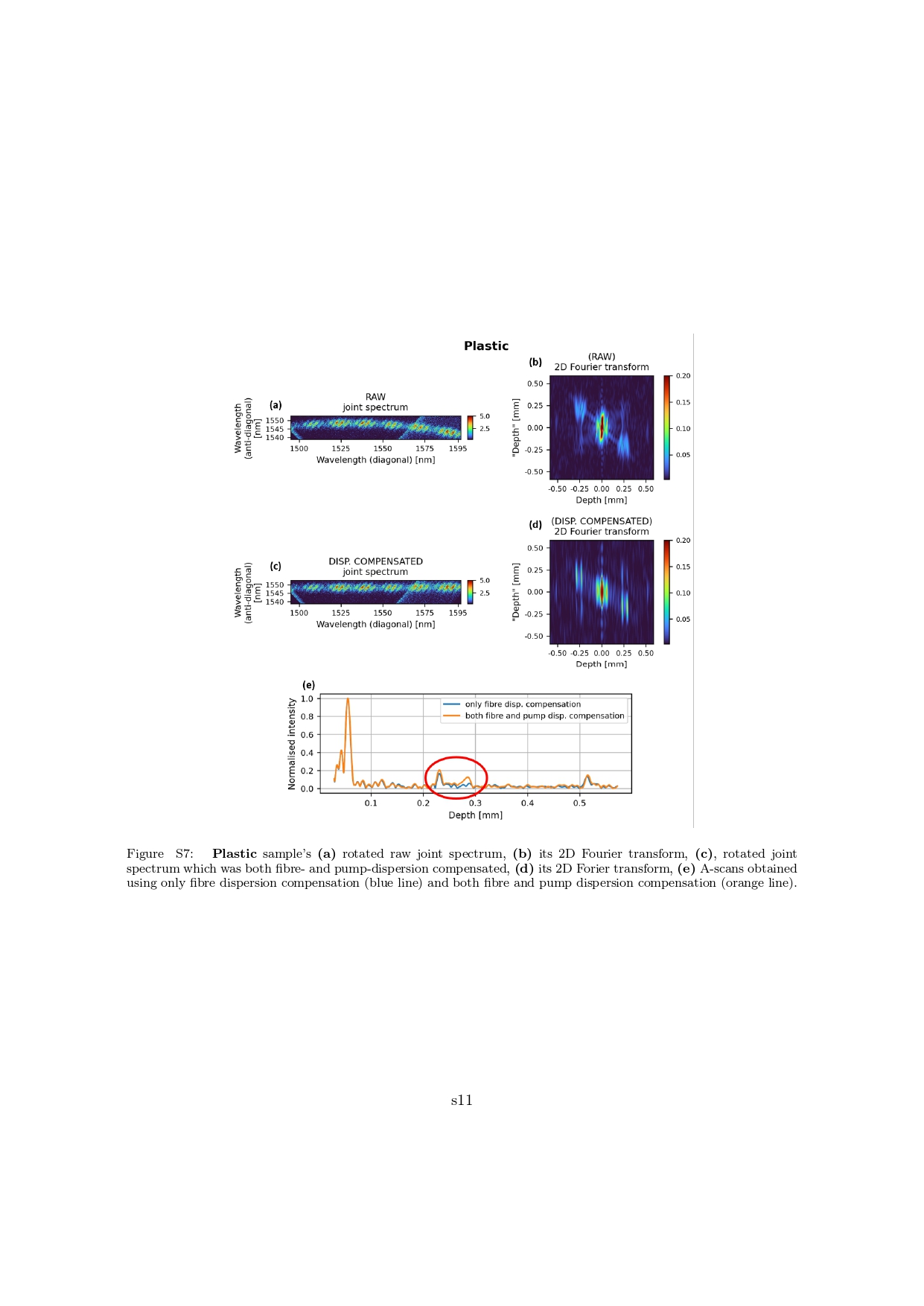}
\end{figure*}

\begin{figure*}[htb]
    %\centering
    \includegraphics[clip=true, trim = 2cm 2cm 2cm 2cm, width=0.95\textwidth]{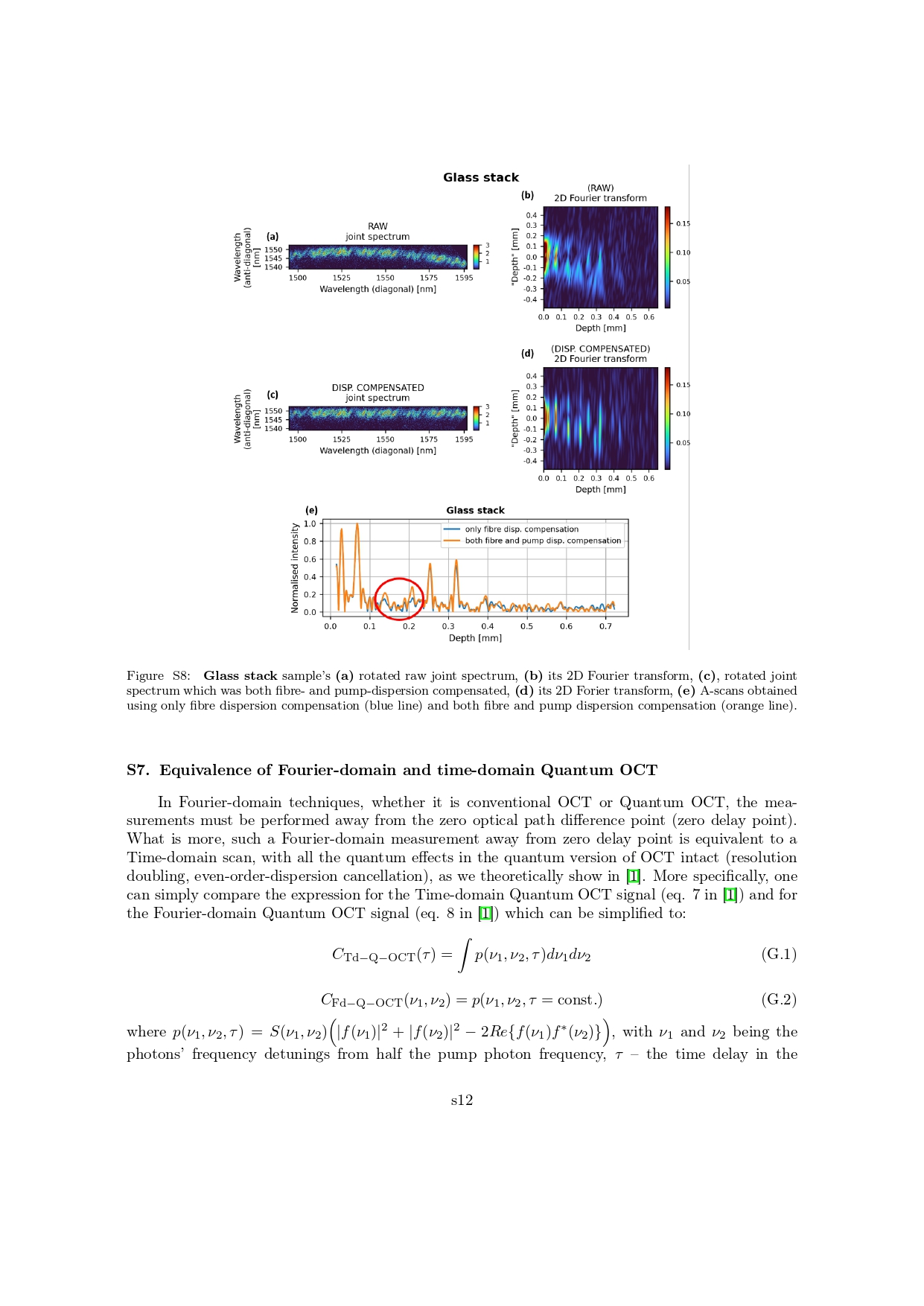}
\end{figure*}

\begin{figure*}[htb]
    %\centering
    \includegraphics[clip=true, trim = 2cm 2cm 2cm 2cm, width=0.95\textwidth]{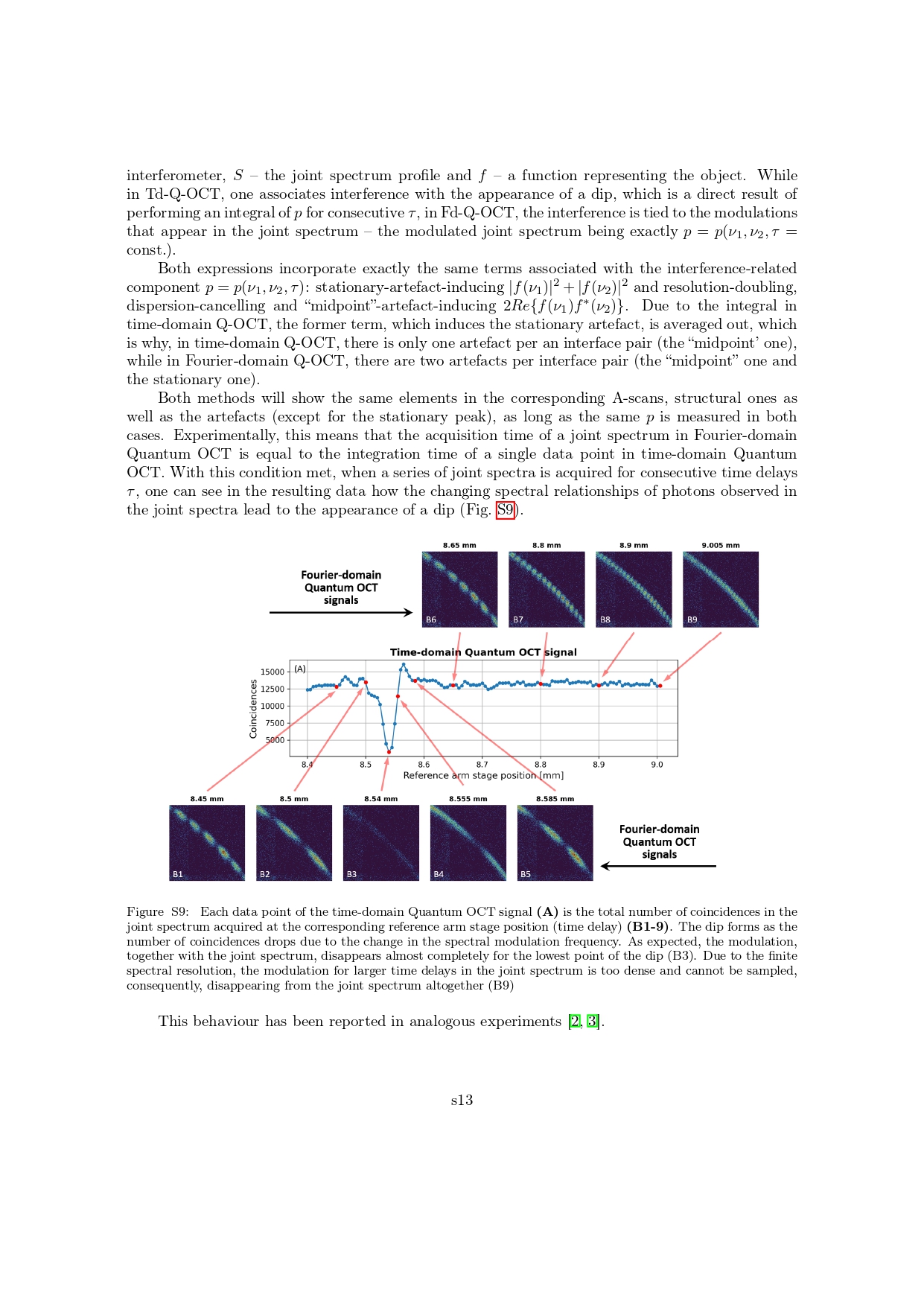}
\end{figure*}

\begin{figure*}[htb]
    %\centering
    \includegraphics[clip=true, trim = 2cm 2cm 2cm 2cm, width=0.95\textwidth]{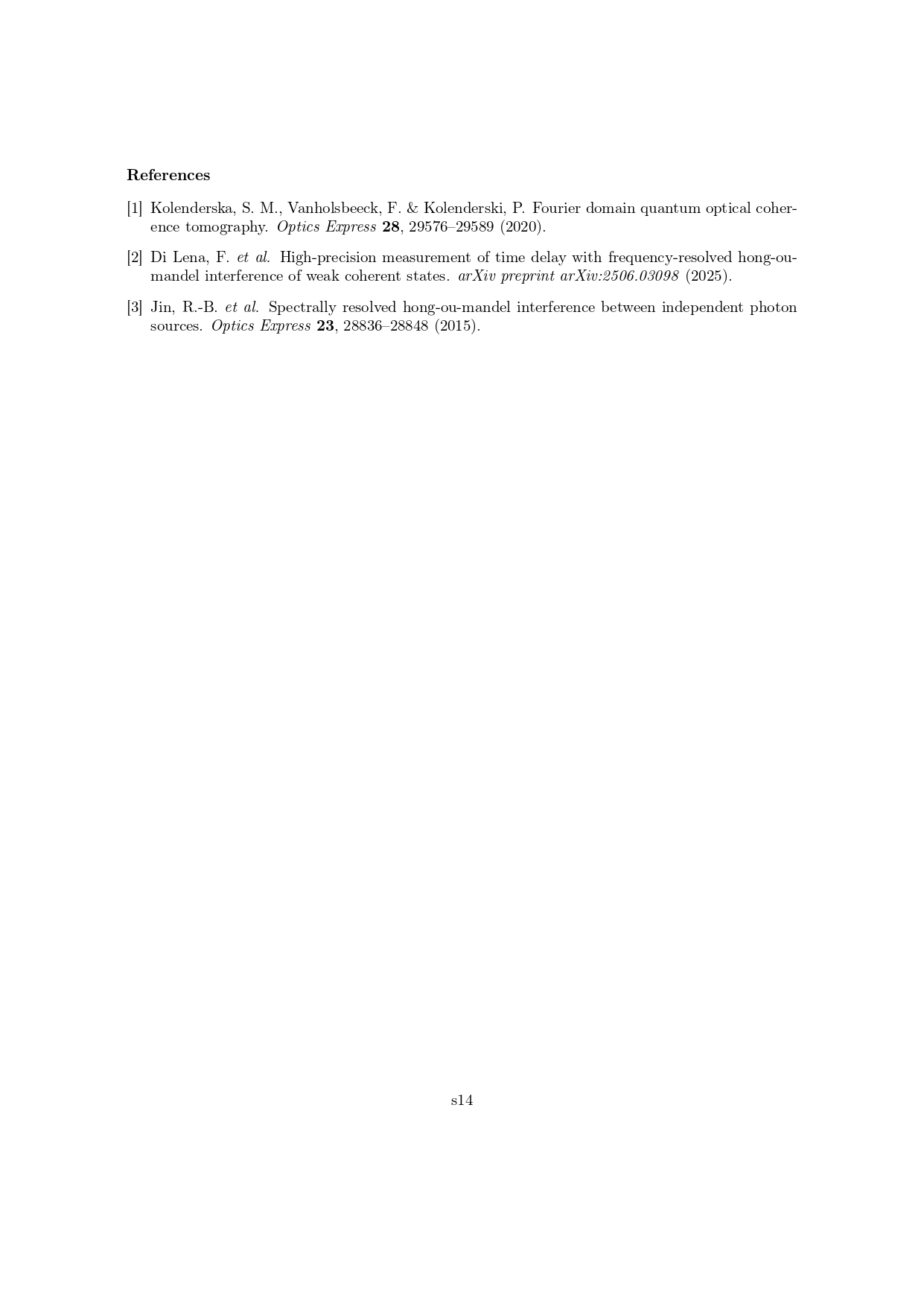}
\end{figure*}

\end{document}